\documentclass[journal,onecolumn]{IEEEtran}%
\usepackage{eurosym}
\usepackage{amsmath}
\usepackage{amsthm}
\usepackage{amsfonts}
\usepackage{amssymb}
\usepackage{graphicx}
\usepackage[usenames,dvipsnames]{color}
\usepackage{booktabs}%
\providecommand{\U}[1]{\protect\rule{.1in}{.1in}}

\theoremstyle{plain}

\theoremstyle{definition}

\newlength{\drop}

\newcommand*{\titleUL}{\begingroup
\drop=0.1\textheight
\vspace*{0.5\drop}
\begin{center}
{\LARGE\textsc{$\quad$}}\\[0.5\drop]
{{\LARGE IEEE Signal Processing Magazine} \\ \vspace*{0.4cm} -- accepted for publication --}\\[\drop]
\rule{\textwidth}{1pt}\par
\vspace{0.5\baselineskip}
{\huge\bfseries The PAPR Problem in OFDM Transmission: \\ \vspace*{0.4cm} New Directions for a Long-Lasting Problem
}\\[0.5\baselineskip]
\rule{\textwidth}{1pt}\par
\vfill
{\Large\textsc{Gerhard Wunder$^1$, Robert F.H. Fischer$^2$, Holger Boche$^3$, \\ Simon Litsyn$^4$, Jong-Seon No$^5$}}
\vfill
$^1$Technische Unversit{\"a}t Berlin, Germany\footnote{Gerhard Wunder is also with the Fraunhofer Heinrich-Hertz-Institut, Berlin.} (\textit{gerhard.wunder@hhi.fraunhofer.de})\\
$^2$Unversit{\"a}t Ulm, Germany (\textit{robert.fischer@uni-ulm.de})\\
$^3$Technische Unversit{\"a}t M{\"u}nchen, Germany (\textit{boche@tum.de})\\
$^4$Tel Aviv University, Israel (\textit{litsyn@eng.tau.ac.il})\\
$^5$Seoul National University, Korea (\textit{jsno@snu.ac.kr})
\vfill
\today
\vfill
\textbf{Keywords}: energy efficiency, RF components, high power amplifier, HPA, peak-to-average power ratio, PAPR, PMEPR, crest-factor, large deviations, derandomization, coding, Banach space geometry, compressed sensing
\vfill
{\itshape \copyright 2012 IEEE. Personal use of this material is permitted. Permission from IEEE must be obtained for all other 
uses, in any current or future media, including reprinting/republishing this material for advertising or promotional purposes, creating new collective works, for resale or redistribution to servers or lists, or reuse of any 
copyrighted component of this work in other works.}
\end{center}
\endgroup}

\newenvironment{tit}{
\title{The PAPR Problem in OFDM Transmission: \\ New Directions for a Long-Lasting Problem}
\author{
\begin{minipage}{.3\linewidth}
\centering \small
Gerhard Wunder\\
Heinrich-Hertz-Institut Berlin, Germany\\
gerhard.wunder@hhi.fraunhofer.de
\end{minipage}
\begin{minipage}{.3\linewidth}
\centering \small
Robert F.H. Fischer\\
Unversit{\"a}t Ulm, Germany\\
robert.fischer@uni-ulm.de
\end{minipage}
\begin{minipage}{.35\linewidth}
\centering \small
Holger Boche\\
Technische Unversit{\"a}t M{\"u}nchen, Germany\\
boche@tum.de
\end{minipage}
\\
\begin{minipage}{.4\linewidth}
\centering \small
Simon Litsyn\\
Tel Aviv University, Israel\\
litsyn@eng.tau.ac.il
\end{minipage}
\begin{minipage}{.4\linewidth}
\centering \small
Jong-Seon No\\
Seoul National University, Korea\\
jsno@snu.ac.kr
\end{minipage}
}
}
\\
\begin{document}
\pagestyle{empty} 
\titleUL
\tit
\thispagestyle{empty}

\addtolength{\baselineskip}{0.1cm}

\section{Energy efficiency in mobile communication networks: A driving source
for innovation}

Energy efficiency particularly matters in future mobile communications
networks. Key driving factor is the growing energy cost of network operation
which can make up as much as 50\% of the total operational cost nowadays
\cite{correia_10}. In the context of \emph{green information and communication
technology} (ICT) this has led to many global initiatives such as the
\emph{Green Touch} consortium\footnote{see, e.g., the webpage:
http://www.greentouch.org}.

A major source for reducing energy costs is to increase the efficiency of the
high power amplifier (HPA) in the radio frequency (RF) front end of the base
stations \cite{correia_10}. However, efficiency of the HPA is directly related
to the \emph{peak-to-average power ratio} (PAPR) of the input signal. The
problem especially becomes serious in orthogonal frequency-division
multiplexing (OFDM) multicarrier transmission which is applied in many
important wireless standards such as the 3GPP Long Term Evolution Advanced
(LTE-A). In the sequel of this article we refer to it simply as the \emph{PAPR
problem}. The PAPR problem still prevents OFDM from being adopted in the
uplink of mobile communication standards, and, besides from power efficiency,
it can also place severe constraints on output power and therefore coverage in
the downlink.

In the past, there have been many efforts to deal with the PAPR problem
resulting in numerous papers and several overview articles, e.g.,
\cite{litsyn_07,han_05,tellado_99_phd}. However, with the upcoming of novel
systems, new challenges emerge which have been rarely addressed so far: 1.)
the envisioned boost in network energy efficiency (e.g. at least by a factor
of 1000 in the Green Touch consortium) will tighten the requirements on
component level so that the efficiency gap with respect to single-carrier
transmission must considerably diminish 2.) multiple-input/multiple-output
(MIMO) multiplicate the problem due to simultaneously control of parallel
transmit signals particularly when considering a huge number of transmit
antennas 3.) multiuser (MU) (and multipoint) systems put additional side
constraints on the parallel transmit signals which are difficult to implement
on top of conventional approaches. Furthermore, many of the existing methods
are not either compatible with relevant standards and/or their prospective
performance capabilities are not satisfactory. Yet, it is quite safe to say
that no standard solution is available.

In this article, we will argue that, in the light of these challenges, the
PAPR metric itself has to be carefully reviewed within a much broader scope
overthrowing some of the common understanding and results. New metrics become
more and more important since they enable the system designer to precisely
adjust the algorithms to meet some given performance indicator. It is expected
that such design approach will no longer be treated like an isolated problem
on physical layer but will affect the design parameters on higher layers as
well (e.g. resource allocation). For example, it has been discussed in
\cite{correia_10} that from a ICT perspective the system throughput should be
related to input power rather than output power. In order to capture this
paradigm on HPA power efficiency level, different metrics are currently used
such as \emph{total degradation}, \emph{average distortion power} and others.
However, with respect to algorithm design all these metrics are solely
reflected by the standard PAPR figure of merit. This argument can also be
extended to other situations: it has been recently shown in
\cite{wunder_11_isit} that, if the only concern is average distortion power
(instead of peak power), then a much less conservative design is possible
compared to conventional design rules in OFDM transmission. Remarkably, such
performance limits can be efficiently achieved using \emph{derandomization}
algorithms establishing therefore a new powerful tool within the context of
the PAPR problem. It is a major aim of this article to review and collect
exactly those elements in the current literature of which we believe represent
the core of a more general theory.

Besides this point of view, it is interesting to apply new signal processing
and mathematical concepts to OFDM. \emph{Compressed sensing}
\cite{Donoho1,Candes1} is a new framework capturing sparsity in signals beyond
Shannon's sampling theorem and has attracted a lot of attention in recent
years. It is based on the observation that a small number of linear
projections (measurements) for a sparse signal contain enough information for
its recovery. Compressed sensing can be applied to the PAPR problem because
sparsity frequently appears in the (clipped) OFDM signals. There are currently
many research efforts in this direction but some challenges still remain, such
as the degraded recovery performance in noisy environment. The adoption of
related mathematical concepts such as \emph{Banach space geometry}
\cite{Kashin,Pinkus} complement this discussion. It is outlined that these
theoretically deeply rooted concepts can help to understand some of the
fundamental limits as well as to develop new algorithmic solutions for the
PAPR problem.

Summarizing, there is a clear need for a fresh look on the PAPR problem under
the general umbrella of the metrics theme discussed before which will open up
new research strands not yet explored. In this article we are going to address
and discuss some of the fundamentals, challenges, latest trends, and potential
solutions which originate from this perspective and which, we believe, are
important to come to an innovative breakthrough for this long-lasting problem.

\subsection{Outline and some notations}

The outline of the article is as follows: first we motivate and discuss
alternative metrics and corresponding methodology for the PAPR problem and
present several examples. Then, we propose an appropriate theoretical
framework and unified algorithm design principles for these new paradigms by
introducing the derandomization principle. In this context, we outline
specific challenges imposed by MIMO and MU MIMO systems. Next, we discuss
capacity issues which establish fundamental limits. Finally, we discuss some
of the future directions the authors believe are the foundations or at least
components of emerging solutions.

We recall the following standard notations: the frequency-domain OFDM symbol
for each antenna ($N_{t}$ in total) consists of $N$ subcarriers. The
multiplexed transmit symbols $C_{m,n}$ (carrying information and/or control
data) are drawn from some common QAM/PSK signal constellation and collected in
the \emph{space-frequency codeword}
{${\mathchoice{\mbox{\boldmath$\displaystyle C$}}                      {\mbox{\boldmath$\textstyle C$}}                      {\mbox{\boldmath$\scriptstyle C$}}                      {\mbox{\boldmath$\scriptscriptstyle C$}}}%
$}%
$:=[{{\mathchoice{\mbox{\boldmath$\displaystyle C$}}                      {\mbox{\boldmath$\textstyle C$}}                      {\mbox{\boldmath$\scriptstyle C$}}                      {\mbox{\boldmath$\scriptscriptstyle C$}}}%
}_{1}%
,...,{{\mathchoice{\mbox{\boldmath$\displaystyle C$}}                      {\mbox{\boldmath$\textstyle C$}}                      {\mbox{\boldmath$\scriptstyle C$}}                      {\mbox{\boldmath$\scriptscriptstyle C$}}}%
}_{N_{t}}]$ where
{${\mathchoice{\mbox{\boldmath$\displaystyle C$}}                      {\mbox{\boldmath$\textstyle C$}}                      {\mbox{\boldmath$\scriptstyle C$}}                      {\mbox{\boldmath$\scriptscriptstyle C$}}}%
$}$_{m}:=[C_{m,1},...,C_{m,N}^{T}]$ is the transmit sequence of antenna $m$.
In case of a single antenna we write $C_{n}:=C_{1,n}$ and, correspondingly,
{${\mathchoice{\mbox{\boldmath$\displaystyle C$}}                      {\mbox{\boldmath$\textstyle C$}}                      {\mbox{\boldmath$\scriptstyle C$}}                      {\mbox{\boldmath$\scriptscriptstyle C$}}}%
$}%
$:={{\mathchoice{\mbox{\boldmath$\displaystyle C$}}                      {\mbox{\boldmath$\textstyle C$}}                      {\mbox{\boldmath$\scriptstyle C$}}                      {\mbox{\boldmath$\scriptscriptstyle C$}}}%
}_{1}=[C_{1},...,C_{m}]^{T}$. Given the IDFT matrix
{${\mathchoice{\mbox{\boldmath$\displaystyle F$}}                      {\mbox{\boldmath$\textstyle F$}}                      {\mbox{\boldmath$\scriptstyle F$}}                      {\mbox{\boldmath$\scriptscriptstyle F$}}}%
$}$:=[\mathrm{e}^{\mathrm{j}2\pi kn/(IN)}]_{0\leq k<IN,0\leq n<N}$ the
$I$-times oversampled discrete-time OFDM transmit symbols in the equivalent
complex baseband at antenna $m$ are given by
{${\mathchoice{\mbox{\boldmath$\displaystyle s$}}                      {\mbox{\boldmath$\textstyle s$}}                      {\mbox{\boldmath$\scriptstyle s$}}                      {\mbox{\boldmath$\scriptscriptstyle s$}}}%
$}$_{m}%
={{\mathchoice{\mbox{\boldmath$\displaystyle F$}}                      {\mbox{\boldmath$\textstyle F$}}                      {\mbox{\boldmath$\scriptstyle F$}}                      {\mbox{\boldmath$\scriptscriptstyle F$}}\mathchoice{\mbox{\boldmath$\displaystyle C$}}                      {\mbox{\boldmath$\textstyle C$}}                      {\mbox{\boldmath$\scriptstyle C$}}                      {\mbox{\boldmath$\scriptscriptstyle C$}}}%
}_{m}$. The average power of this signal might be normalized to one. We define
the PAPR of the transmit signal at antenna $m$ as
\begin{equation}
\mathrm{PAPR}\left(
{{\mathchoice{\mbox{\boldmath$\displaystyle s$}} {\mbox{\boldmath$\textstyle s$}} {\mbox{\boldmath$\scriptstyle s$}} {\mbox{\boldmath$\scriptscriptstyle s$}}}%
}_{m}\right)  :=\left\Vert
{{\mathchoice{\mbox{\boldmath$\displaystyle s$}} {\mbox{\boldmath$\textstyle s$}} {\mbox{\boldmath$\scriptstyle s$}} {\mbox{\boldmath$\scriptscriptstyle s$}}}%
}_{m}\right\Vert _{\infty}^{2}.
\end{equation}
\emph{Comment on oversampling}: Please note that PAPR of the continuous-time
passband signal differs roughly by 3dB. Clearly, there is also still some
overshooting between the samples but due to sufficiently high oversampling the
effect is negligible. The trade off between overshooting and oversampling is
one of the few subproblems in OFDM transmission that is well understood. The
best known results which hold even in the strict band-limited case are given
in \cite{wunder_03_sig} where overshooting is proved to be below $1/\cos
(\frac{\pi}{2I})$.

\section{The design challenge}

\label{sec:design}

In OFDM transmission many subcarriers (constructively or destructively) add up
at a time which causes large fluctuations of the signal envelope; a
transmission which is free from any distortion requires linear operation of
HPA over a range $N$ times the average power. As practical values of
subcarriers are large this high dynamics affords HPA operation well below
saturation so that most of the supply power is wasted with deleterious effect
on either battery life time in mobile applications (uplink) or energy cost of
network operation (downlink). In practice, these values are not tolerable and
from a technology viewpoint it is also challenging to provide a large linear
range. Hence, the HPA output signal is inevitably cut off at some point
relative to the average power (\emph{clipping level}) leading to in-band
distortion in the form of intermodulation terms and spectral regrowth into
adjacent channels. The effect is illustrated in Fig. \ref{fig:time_signal}
where the distorted OFDM signal and corresponding impact on the signal points
are depicted.

The PAPR problem brings up several challenges for the system designer: one
challenge is to adjust HPA design parameters (HPA backoff, digital
predistortion) in some specific way so that power efficiency is traded against
nonlinear distortion which effects the data transmission on a global scale. To
capture this trade off by a suitable metric on the level of the HPA is far
from clear yet. Special HPA architectures at component level such as Doherty
\cite{raab02} and others can help to improve on this trade off. We also
mention that other design constraints such as costs might prevent specific
architectures \cite{fettweis_07}.

A second challenge is to process the baseband signal by peak power reduction
algorithms in such a way that the key figures of merit in the before mentioned
trade off are improved. This alternating procedure makes apparent that the
PAPR problem involves joint optimization of HPA, predistortion and signal
processing unit. This interplay has only been marginally addressed so far let
alone in the context of multiuser systems equipped with multiple antennas such
as LTE-A.

In the following we discuss some potential metrics that can be used in the optimization.

\section{The right paradigm? Alternative metrics for PAPR}

\label{sec:alternative}

Classically, in OFDM transmission the PAPR of the transmit signal is analyzed
and minimized by applying transmitter-side algorithms. Meanwhile it has been
recognized that it may be reasonable to study other parameters as well.
Especially when aiming at minimizing the energy consumption of the transmitter
including the analog front end or when operating low-cost, low-precision power
amplifier---sometimes referred to as \textquotedblleft dirty
RF\textquotedblright\ \cite{fettweis_07}---potentially other signal properties
need to be controlled.

Let us present some illustrative example first. Suppose, we are interested in
the clipped energy instead of the PAPR (we give some justification in terms of
capacity below for this). Naturally, since the total energy is approximately
one the clipped energy is finite as well but when $N$ increases the required
clipping level for asymptotically zero clipping energy might be of interest
for design purposes. Clearly, no clipping at all is trivially sufficient but,
surprisingly, it is actually \emph{not} necessary: it is proved in
\cite{wunder_11_isit} that clipping level can be adjusted along the $\log
\log\left(  N\right)  $ law so that it is practically almost constant. This
stands in clear contrast to the $\log\left(  N\right)  $ PAPR scaling
discussed in Sec. \ref{sec:ldp}.

Subsequently, some alternative metrics replacing the PAPR value in specific
situations are briefly summarized.

\begin{figure}[ptb]
\begin{center}
\includegraphics[
height=3.5228in,
width=3.2131in
]{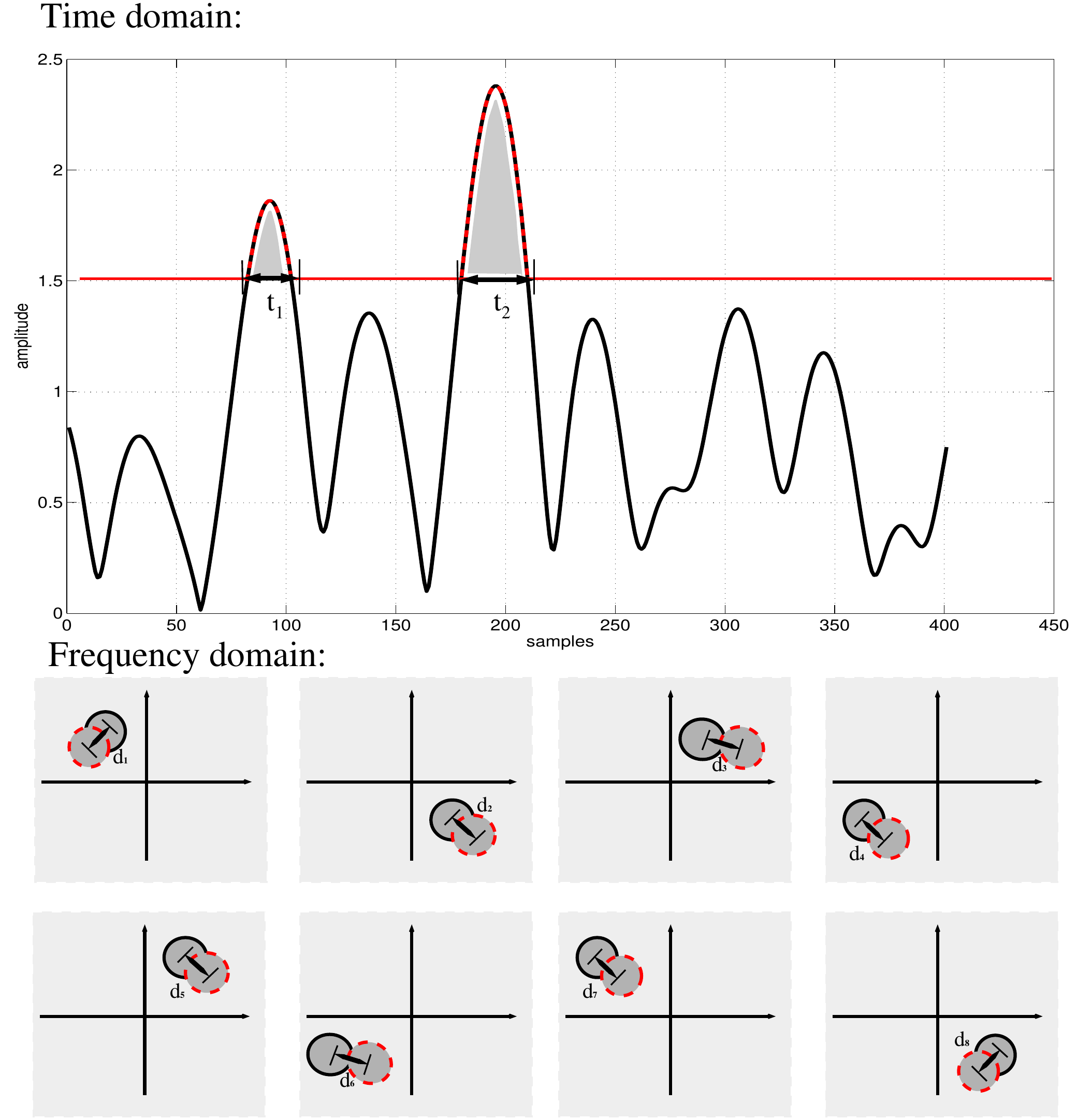}
\end{center}
\caption{Illustration of distorted OFDM signal (in time domain, clipping
level=3.5 dB) and corresponding impact on the subcarrier signalling points (in
frequency domain): the distortion signal is typically a sequence of
\emph{clips} with clip duration $t_{1},t_{2},...$ The SDR metric then relates
the mean useful signal energy to the mean distortion energy while EVM collects
the mean of sum of squared errors $d_{1},d_{2},...$ in the data sequence (due
to the very same time domain distortion). In case of Nyquist-sampling both are
actually equal (subject to a scaling factor) \cite{kotzer_10,Kotzer2011}.}%
\label{fig:time_signal}%
\end{figure}

\begin{itemize}
\item Of course, the \emph{PAPR} has still its justification. As it relates
peak and mean power the PAPR is the adequate metric for quantifying the
required input power backoff of the power amplifier. When using higher-order
modulation per carrier the energy per OFDM frame is no longer constant and
hence average power fluctuates. In this case, the \emph{peak power} is as
suited metric. Both metrics are well-suited measures if purely limiting
effects (modeled, e.g., as soft limiter) should be characterized. Since PAPR
is random, we are also interested in the complementary cumulative distribution
function (CCDF) $F(x)$ and other characteristic figures such as the mean
$\mathbb{E}\left\{  \cdot\right\}  $ etc.

\item Besides looking at the transmit signal itself, in many situations the
impact of the nonlinear power amplifier to system performance is of interest.
One possible approach is to quantify the nonlinear distortions caused by a
particular power amplifier model. The \emph{signal-to-distortion ratio (SDR)}
\cite{ochiai_02_comm,hua_04} or \emph{error vector magnitude (EVM)}
\cite{kotzer_10,Kotzer2011} capture the in-band distortion of the OFDM signal
and are immediately related to error rate of uncoded transmission. Both, SDR
and EVM and their interdependence are illustrated in Fig.
\ref{fig:time_signal}.

\item Within 3GPP, a power amplifier model which cause non-linear distortions
according to the third power of the RF transmit signal has become popular
(so-called cubic polynomial model) \cite{3gpp_04}. In this case, the
\emph{cubic metric (CM)} measuring the mean of time domain sample energy to
the third power is well adapted to the specific scenario
\cite{3gpp_09,siegl_2010}. However, any additional clipping is not included here.

\item CM metric is a special case of the \emph{amplifier oriented metric
(AOM)} defined in \cite{siegl_09,siegl_2010}. AOM measures the mean squared
absolute difference of desired and distorted HPA output. Here, the HPA output
is calculated based upon some model such as the mentioned cubic polynomial
model (or the well-known Rapp model etc.).

\item A much severe problem in communication systems is that nonlinear devices
cause a spectral widening and hence generate out-of-band radiation. In order
not to violate spectral masks imposed by the regulatory body, a metric
quantifying the out-of-band power or the shoulder attenuation is desirable
\cite{siegl_09}. In this field, significant work has to be done.

\item When applying strong channel coding schemes---which nowadays is the
state of the art in OFDM transmission---SDR or EVM are no longer suited
performance characteristics. Instead, the end-to-end \emph{capacity} of the
entire OFDM scheme including the nonlinear devises matters. Unfortunately,
neither the capacity of the continuous-time peak-power limited additive white
Gaussian noise (AWGN) channel itself, nor the capacity of OFDM over such
channels are known. Moreover, when using (as often done) a statistical model
for the behavior of the nonlinear device, only lower bounds on the capacity
are obtained as the statistical dependencies within one OFDM frame are not exploited.

However, recent work \cite{peng_06,fischer_10} indicates that clipped OFDM
performs (almost) the same as unclipped OFDM with signal-to-noise ratio
reduced according to the clipping power loss. Hence, the main source of the
loss are not the introduced distortions or errors but simply the reduced
output power. This, in turn, leads to the conclusion that a suited metric for
capacity maximization is simply the average power of the power amplifier
output signal. In this case, unfortunately, the generation of out-of-band
radiation is not penalized.

\item The symbol error rate (SER) is a related measure which has been directly
applied to peak power control algorithms in \cite{WS10apat}.

\item In future applications, more than a single signal parameter will have to
be controlled. E.g., the capacity should be maximized but at the same time the
out-of-band power should be minimized. Consequently, suited combinations of
metrics capturing the desired trade off are requested.
\end{itemize}

\begin{figure}[t]
\begin{center}
\includegraphics[
height=5.9cm,
width=0.6\columnwidth
]{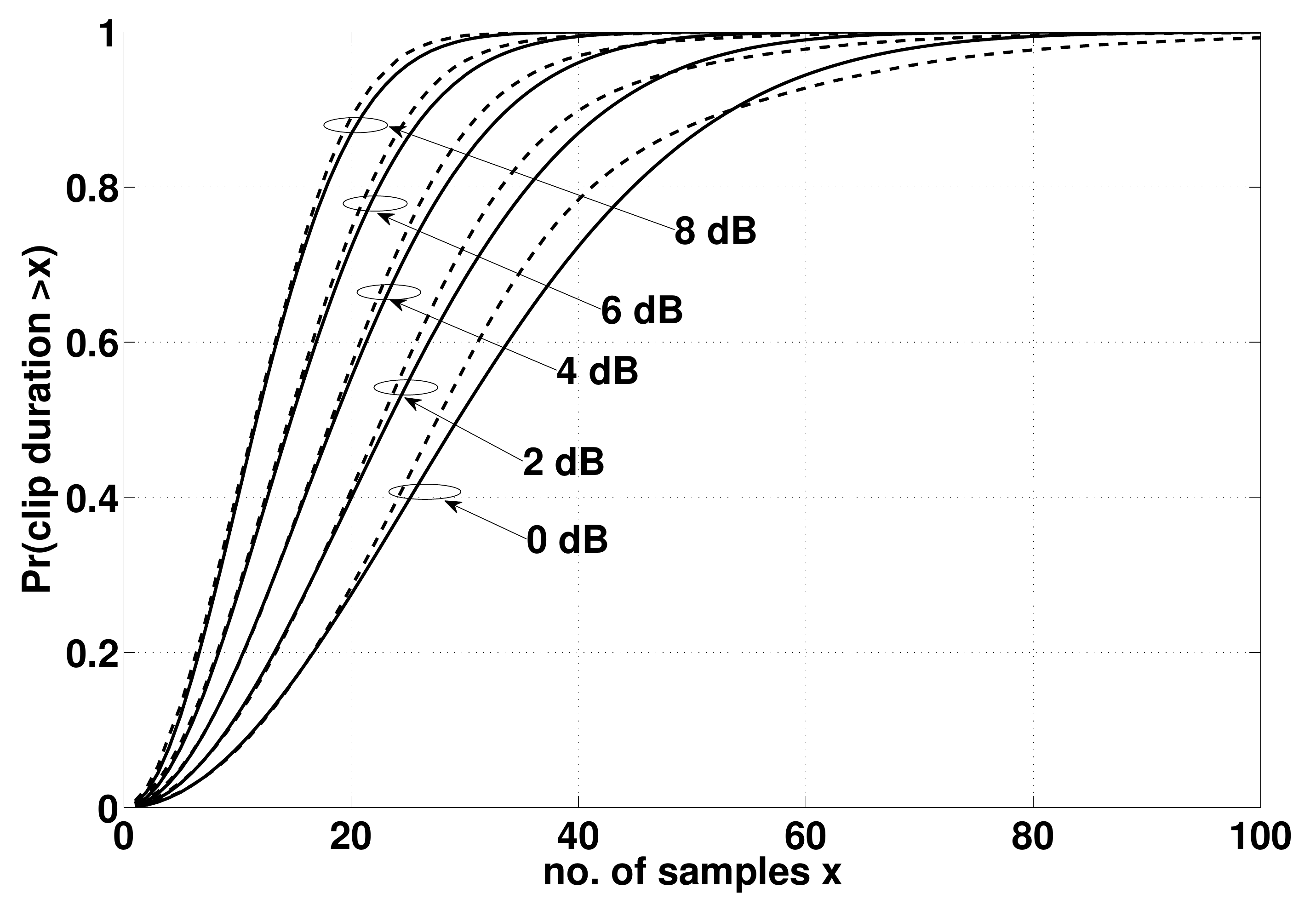}
\end{center}
\caption{Palm distribution of clip duration of an OFDM signal with $N=2048$
and different clipping levels in [dB]. Note that the theory (dotted) is valid
only for large clipping levels but fits relatively well also for low clipping
levels.}%
\label{fig:ccdf}%
\end{figure}

\emph{Comment on Gaussian approximation}: Noteworthy, many metrics have been
analyzed in the past with the help of the Gaussian approximation. This,
however, is not in all cases a feasible path. It is true that as $N$ gets
large the finite-dimensional distributions converge and since the signals are
band-limited also the process itself. However, we mention that it is a local
property, valid only for any finite interval. For example in Fig.
\ref{fig:ccdf} the empirical CCDF of the clip duration (see the illustration
in Fig. \ref{fig:time_signal}) is shown for OFDM signals and compared to a
widely unknown result for envelopes of Gaussian processes (and their Hilbert
transforms) for large clipping levels \cite{Belyaev69}\footnote{To be
specific, Ref. \cite{Belyaev69} derives the so-called \emph{Palm distribution}
of the clip duration which describes the statistical average after an
upcrossing of the Gaussian process.}. It is seen that simulation and analysis
match very well \cite{WS10apat}. On the other hand, metrics such as EVM
\cite{Kotzer2011} and SER \cite{Henkel2004} have been shown not to match well.

\section{Approaching the $\log\left(  N\right)  $ barrier: Derandomization}

\label{sec:ldp}

In this section we discuss several fundamental principles for the peak power
control problem. We believe that all of them actually connect to a broader
theory general enough to capture alternative metrics as well and will open the
door for new, provably more efficient algorithms.

\subsection{The LDP}

\label{sec:ldp_ldp}

By analyzing PAPR of multicarrier signals one faces a fundamental barrier
which to overcome seems quite challenging: the $\log\left(  N\right)  $
barrier (recall: $N$ is the number of subcarriers). In fact, it is an exercise
in large deviations to show that multicarrier signals with statistically
independent subcarriers have PAPR of $\log\left(  N\right)  $ in a
probabilistic sense \cite{ochiai_01_comm,dinur_01_comm,wunder_03_inf}. This
means that with very high probability the PAPR lies in an open, arbitrarily
small interval containing $\log\left(  N\right)  $: this is what we call the
\emph{large deviation principle} or in short \emph{LDP}.

Implicitly, LDP affects the performance of many peak power control schemes.
The LDP has been known since long in the context of random polynomials but in
the OFDM context the most general form is due to \cite{litsyn_06_inf} where it
is shown that as long as $N$ is large enough and the subcarriers are
independent that the following inequality is true:%
\begin{gather}
\Pr\left\{  \left\vert \sqrt{\mathrm{PAPR}}-\sqrt{\mathrm{\log}\left(
N\right)  }\right\vert >\gamma\frac{\mathrm{\log}\left[  \mathrm{\log}\left(
N\right)  \right]  }{\sqrt{\mathrm{\log}\left(  N\right)  }}\right\}
\nonumber\\
\leq\frac{1}{\left[  \log\left(  N\right)  \right]  ^{2\gamma-\frac{1}{2}}}
\label{eqn:ldp_sqrt}%
\end{gather}
Here, {$\gamma>\frac{1}{4}$ is some design parameter which trades off
probability decay over deviation from }$\mathrm{\log}\left(  N\right)  ${. }
While the analysis is tricky when it comes to show that PAPR is not below
$\mathrm{\log}\left(  N\right)  $ with high probability, it is a surprisingly
easy task to show the converse: standard inequalities (such as Chernoff
bounds) or any other Markov-style bound do the job. Some can be exploited for
algorithm design.

The inequality states that PAPR concentrates more and more around the value
$\log\left(  N\right)  $ which establishes therefore an important theoretical
scaling law. The proof is technical but the result might be surprising since
1)\ the factor before the logarithmic term is exactly unity and 2) the scaling
law differs from the well-known law of iterated logarithm which would suggest
only doubly logarithmic scaling.

\begin{figure}[ptb]
\begin{center}
\includegraphics[
height=2.5166in,
width=2.8in
]{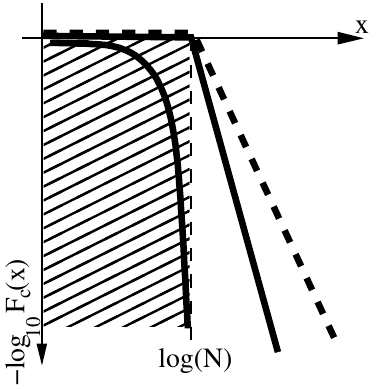}
\end{center}
\caption{Illustration of the virtue of the LDP: [from right to left] CCDF of
PAPR of a.) uncoded data b.) SLM and c.) derandomization.}%
\label{fig:spm_log}%
\end{figure}

The LDP contains some valuable illustrative aspects which we are going to
reveal now. The LDP in eqn. (\ref{eqn:ldp_sqrt}) is somewhat unaccessible and
shall be rewritten in the more convenient form:
\begin{equation}
\log\left(  F_{c}\left(  x\right)  \right)  =\left[  \log\left(  N\right)
+O\left(  \log\left[  \log\left(  N\right)  \right]  \right)  -x\right]  ^{-}%
\end{equation}
where we used the order notation $O\left(  \cdot\right)  $ and the definition
$\left[  x\right]  ^{-}:=\min\left(  0,x\right)  $. Disregarding the order
term $O\left(  \log\left[  \log\left(  N\right)  \right]  \right)  $ we have
the interpretation that the probability decreases linearly on a logarithmic
scale from some cut-off point $\log\left(  N\right)  $ on which is illustrated
in Fig. \ref{fig:spm_log}. The proximity to filter design terminology is
intended and it makes obviously sense to speak of a pass band and a stop band
in the figure. Comparing this to the standard analysis where statistical
independence and Nyquist sampling is assumed gives%
\begin{equation}
\log\left(  F_{c}\left(  x\right)  \right)  =\left[  \log\left(  N\right)
-x\right]  ^{-}%
\end{equation}
where the order term is missing. Hence, we conclude that a careful
non-Gaussian analysis for continuous-time OFDM signals entails an error of at
most $O\left(  \log\left[  \log\left(  N\right)  \right]  \right)  $.

The LDP is very useful for assessing the performance of peak-power control
schemes. Before we show this we might ask why this concentration happens? Let
$C_{1},C_{2},...,C_{N}$ be a (data) sequence of independent random variables;
when estimating PAPR without a priori information the expectation is the best
possible choice. Using successive knowledge of already fixed data we have the
following estimations:%
\begin{align}
y_{0}  &  =\mathbb{E}\left\{  \mathrm{PAPR}\left(
{{\mathchoice{\mbox{\boldmath$\displaystyle s$}} {\mbox{\boldmath$\textstyle s$}} {\mbox{\boldmath$\scriptstyle s$}} {\mbox{\boldmath$\scriptscriptstyle s$}}}%
}\right)  \right\} \label{eqn:ldp_b}\\
y_{1}  &  =\mathbb{E}\left\{  \mathrm{PAPR}\left(
{{\mathchoice{\mbox{\boldmath$\displaystyle s$}} {\mbox{\boldmath$\textstyle s$}} {\mbox{\boldmath$\scriptstyle s$}} {\mbox{\boldmath$\scriptscriptstyle s$}}}%
}\right)  |C_{1}\right\} \\
y_{2}  &  =\mathbb{E}\left\{  \mathrm{PAPR}\left(
{{\mathchoice{\mbox{\boldmath$\displaystyle s$}} {\mbox{\boldmath$\textstyle s$}} {\mbox{\boldmath$\scriptstyle s$}} {\mbox{\boldmath$\scriptscriptstyle s$}}}%
}\right)  |C_{1},C_{2}\right\} \\
&  ...\nonumber\\
y_{N}  &  =\mathbb{E}\left\{  \mathrm{PAPR}\left(
{{\mathchoice{\mbox{\boldmath$\displaystyle s$}} {\mbox{\boldmath$\textstyle s$}} {\mbox{\boldmath$\scriptstyle s$}} {\mbox{\boldmath$\scriptscriptstyle s$}}}%
}\right)  |C_{1},C_{2},...,C_{N}\right\}  \label{eqn:ldp_e}%
\end{align}
It can be shown that this process establishes a \emph{Martingale} with bounded
increments $\left\vert y_{i}-y_{i-1}\right\vert $ from which it follows (see
\cite{sason11}) measure concentration of the PAPR around its average via the
\emph{Azuma-Hoeffding} inequality or \emph{McDiarmid}'s inequality. Another
approach used in \cite{sason11} for proving measure concentration of the PAPR
around its median is based on the convex-hull distance inequality of
\emph{Talagrand}. The tails of the concentration inequalities are even
exponential then. Let us now apply the LDP within the context of peak power control.

\subsection{Multiple signal representation and partitioning}

\label{sec:ldp_part}

\begin{figure}[ptb]
\begin{center}
\includegraphics[
height=5.2324cm,
width=7.8595cm
]{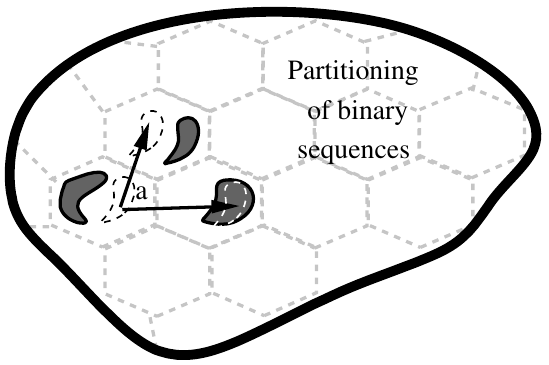}
\end{center}
\caption{A general model for peak power control: each cell contains the same
information; sets with large PAPR (dotted areas) are mapped to sets where PAPR
is below some threshold (grey-shaded areas).}%
\label{fig:spm_peakred}%
\end{figure}

The basic principle for most of the peak power control schemes is
\emph{multiple signal representation} which roots in the classical methods
\emph{selected mapping} (SLM) \cite{baeuml_96_el} and \emph{partial transmit
sequences} (PTS) \cite{mueller_97_el,alavi_05_cml}. The idea is simple:
instead of transmitting the original OFDM data frame multiple redundant
candidates are generated and the \textquotedblleft best\textquotedblright%
\ candidate is singled out for transmission. By using suitable transforms or
mappings the main goal is to achieve statistical independence between the
candidates' metrics. Clearly, instead of PAPR, also alternative metrics can be
used in the selection process \cite{siegl_09,siegl_2010}. SLM and PTS are
similar, the difference between SLM and PTS is that the mappings are applied
to a subset of the data frame.

For SLM many transforms have been proposed: in the original approach the data
frame is element-wise multiplied by random phases; other popular approaches
include binary random scrambling and permutation of the data (see ref. in
\cite{siegl_2010}; here is it also where side transmission is discussed).
Similar for PTS, random phases have been used as well. While typically a full
search is carried out efficient algorithms to find the phases have been
proposed. An exhaustive list can be found in \cite{litsyn_07}.

SLM can be analyzed within the context of LDP. The transforms define $U$
alternatives each assumed with independent PAPR. Clearly this independence
assumption is crucial: it might be argued that it has to hold for the PAPR
only but the model clearly fails when the number of alternatives is large. By
exploiting the LDP we have simply then%
\begin{equation}
\log\left(  F_{c}\left(  x\right)  \right)  =U\left[  \log\left(  N\right)
-x\right]  ^{-}%
\end{equation}
so that the decay is $U$ times faster as depicted in Fig. \ref{fig:spm_log}. A
similar analysis can be carried out when the selection is done directed or
over extended time \cite{siegl_2010}. Note that in principle PTS can be
analyzed as well; however since the transform is on subsets of the data the
independence assumption is far more critical. Another main problem so far is
that side information is treated separately and \emph{not} within the same
communication model.

A better model is complete \emph{partitioning} of the set of transmit
sequences. The idea is illustrated in Fig. \ref{fig:spm_peakred}. Suppose that
the transmit sequence belong to some set which is partitioned into many cells
all of them containing the same information. Note that if the actual cell
selection is required at the receiver for decoding, side information is
generated. This side information belongs in our general model to the transmit
sequence itself and must be specially protected. This can be done via an
embedded code which is decoded before or after the actual information decoding
procedure \cite{litsyn_07}. Let us mark the subsets where PAPR is below some
threshold: the reasoning is that by the mapping of codewords from one cell
into another, sets with larger PAPR should be mapped to a marked subset by at
least one mapping which will ensure peak power below the threshold. Obviously,
the definition of such a mapping will determine the performance of the scheme.

One of the simplest examples is when the data is over some constellation and
side information is encoded into a sequence of BPSK symbols: each sequence
defines a specific BPSK vector determining the sign vector. Both modified
information and side information sequence define the transmitted codeword.
This method is called \emph{sequence balancing} \cite{litsyn_07_toc}. It is
characteristic for this method that correlation is inserted in the stream by
using suitable binary codes. We will call this the \emph{binary correlation
model}. Noteworthy, if the side information is purely redundant the method
reduces to \emph{tone reservation} \cite{tellado_99_phd}. Moreover, if the
selection defines phase relations between partial sequences then it is a
version of PTS \cite{mueller_97_el}. Related approach is also Trellis shaping
\cite{Nguyen08,litsyn_07}.

Sequence balancing using binary codes can achieve (even though easily
generalized) already a sufficient fraction of the theoretically possible
performance gain: the main required property of the set of binary vectors is
their ability of as many sign changes as possible over any subvector which is
called the \emph{strength of the code} \cite{litsyn_07_toc}. Many binary codes
have this property and are thus suited for this procedure. The strength is
related to the dual distance. It can be shown that if the strength grows as
$\log\left(  N\right)  $ then PAPR is below $\log\left(  N\right)  $ for large
$N$. Unfortunately, similar to SLM and PTS, the number of candidates grows as well.

There are other methods which use partitioning as well such as \emph{tone
injection} \cite{tellado_99_phd} where the constellation is artificially
extended or \emph{translates of codes} \cite{schmidt_10_cl}. Schemes such as
\emph{active constellation extension} \cite{Krongold_Jones_2003} introduce
redundancy as well but can be continuously formulated so that other methods
such as convex optimization can be applied.

All discussed approaches assume to run a full cell selection search which is
too complex in many situations. A better approach is discussed next.

\subsection{Derandomization of choices}

The LDP provides a method to circumvent full search by assuming a suitable
underlying probability model for the cell selection. By derandomizing the cell
selection one can easily devise suitable algorithms guaranteeing a PAPR
reduction very close to the log$\left(  N\right)  $ barrier
\cite{sharif_04_sig,litsyn_06_inf,sharif_09_toc}. The basis algorithm goes
back to Spencer \cite{spencer_94} who called it the \emph{probabilistic
method}.

\begin{figure}[t]
\centering\includegraphics[width=0.6\columnwidth,height=5.9cm]{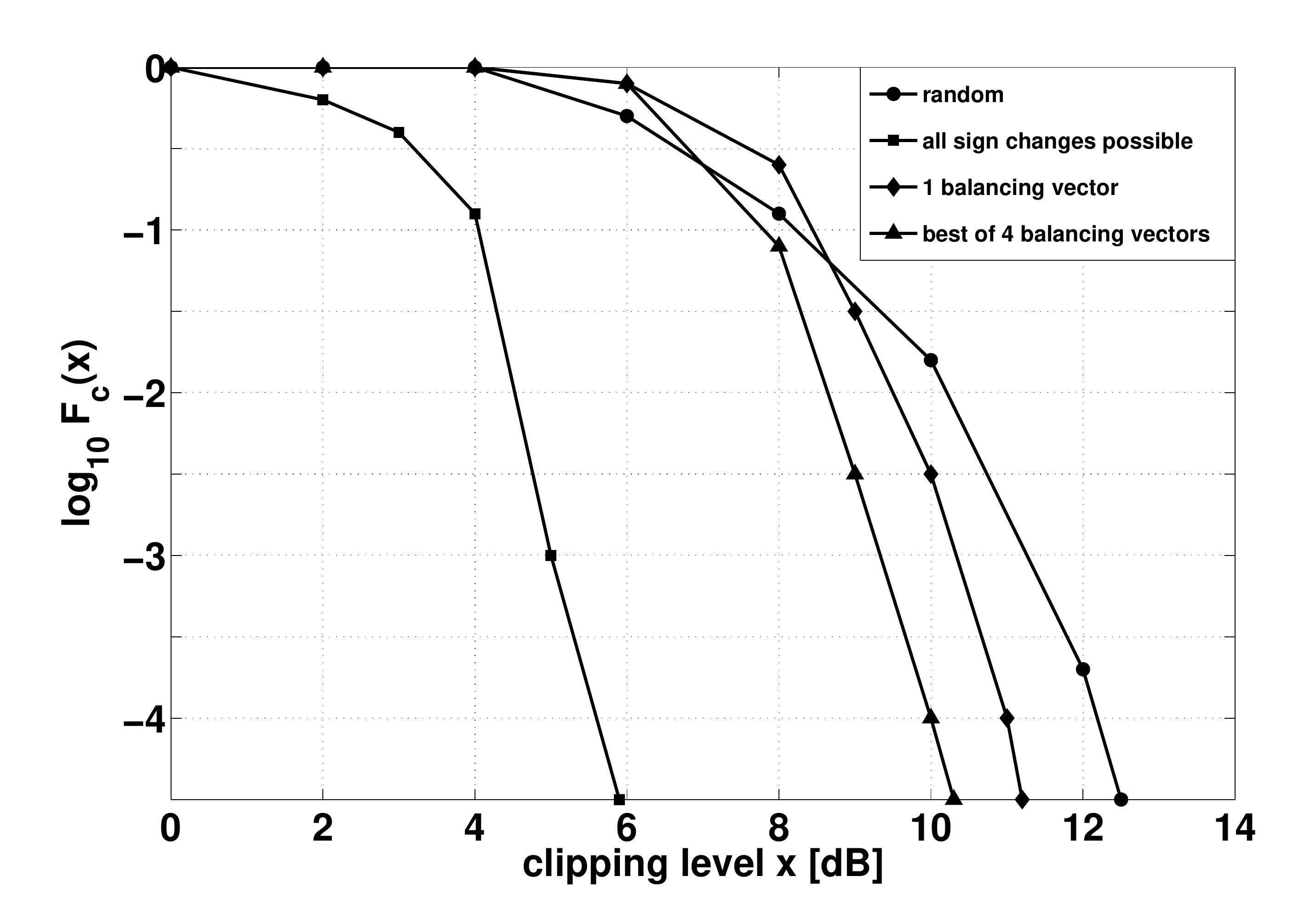}\caption{The
figure compares the sequence balancing method of Sec. \ref{sec:ldp_part} in
terms of the CCDF of PAPR for a 128 subcarrier OFDM system using a.)
derandomization algorithm where all possible subcarrier sign changes are
allowed and b.) codes of given strength with randomly chosen balancing
vectors. For the code a strength 10 dual BCH code with only 18 redundant
subcarrier is applied. Please note that the simulation matches very well the
theory predicting cut off point $\log\left(  128\right)  \approx6.8$ dB.}%
\label{fig:litsyn}%
\end{figure}

The derandomization method is best explained along an example: consider again
the binary correlation model where any possible sign change for some
information sequence
{${\mathchoice{\mbox{\boldmath$\displaystyle C$}}                      {\mbox{\boldmath$\textstyle C$}}                      {\mbox{\boldmath$\scriptstyle C$}}                      {\mbox{\boldmath$\scriptscriptstyle C$}}}%
$ }is allowed. Denote this sign vector by
{${\mathchoice{\mbox{\boldmath$\displaystyle A$}}                      {\mbox{\boldmath$\textstyle A$}}                      {\mbox{\boldmath$\scriptstyle A$}}                      {\mbox{\boldmath$\scriptscriptstyle A$}}:=[}%
A$}$_{0},\ldots,A_{N-1}]$ and the resulting transmit sequence by
{${\mathchoice{\mbox{\boldmath$\displaystyle A$}}                      {\mbox{\boldmath$\textstyle A$}}                      {\mbox{\boldmath$\scriptstyle A$}}                      {\mbox{\boldmath$\scriptscriptstyle A$}}}%
$}$\circ$%
{${\mathchoice{\mbox{\boldmath$\displaystyle C$}}                      {\mbox{\boldmath$\textstyle C$}}                      {\mbox{\boldmath$\scriptstyle C$}}                      {\mbox{\boldmath$\scriptscriptstyle C$}}}%
$} (respectively
{${\mathchoice{\mbox{\boldmath$\displaystyle s$}}                      {\mbox{\boldmath$\textstyle s$}}                      {\mbox{\boldmath$\scriptstyle s$}}                      {\mbox{\boldmath$\scriptscriptstyle s$}}}%
$}%
$_{{{\mathchoice{\mbox{\boldmath$\displaystyle A$}}                      {\mbox{\boldmath$\textstyle A$}}                      {\mbox{\boldmath$\scriptstyle A$}}                      {\mbox{\boldmath$\scriptscriptstyle A$}}}%
}\circ
{{\mathchoice{\mbox{\boldmath$\displaystyle C$}}                      {\mbox{\boldmath$\textstyle C$}}                      {\mbox{\boldmath$\scriptstyle C$}}                      {\mbox{\boldmath$\scriptscriptstyle C$}}}%
}}$). Suppose that all the sign changes happens at random with equal
probability and each sign change is independent. As for the LDP, define the
random variables $y_{i}\left(  A_{0}^{\ast},\ldots,A_{i-1}^{\ast}\right)
:=\mathbb{E}\left\{  \mathrm{PAPR}\left(
{{\mathchoice{\mbox{\boldmath$\displaystyle s$}}                      {\mbox{\boldmath$\textstyle s$}}                      {\mbox{\boldmath$\scriptstyle s$}}                      {\mbox{\boldmath$\scriptscriptstyle s$}}}%
}%
_{{{\mathchoice{\mbox{\boldmath$\displaystyle A$}}                      {\mbox{\boldmath$\textstyle A$}}                      {\mbox{\boldmath$\scriptstyle A$}}                      {\mbox{\boldmath$\scriptscriptstyle A$}}}%
}\circ
{{\mathchoice{\mbox{\boldmath$\displaystyle C$}}                      {\mbox{\boldmath$\textstyle C$}}                      {\mbox{\boldmath$\scriptstyle C$}}                      {\mbox{\boldmath$\scriptscriptstyle C$}}}%
}}\right)  |A_{0}^{\ast},\ldots,A_{i-1}^{\ast}\right\}  $. Then we can mimic
the steps (\ref{eqn:ldp_b})-(\ref{eqn:ldp_e}) and successively reduce
randomness by applying:%
\[
A_{i}^{\ast}:=\underset{a_{i}}{\arg\min}\;y_{i}\left(  A_{0}^{\ast}%
,\ldots,A_{i-1}^{\ast},A_{i}\right)
\]
By the properties of (conditional) expectations%
\[
y_{i}\left(  A_{0}^{\ast},\ldots,A_{i-1}^{\ast},A_{i}^{\ast}\right)  \leq
y_{i}\left(  A_{0}^{\ast},\ldots,A_{i-1}^{\ast}\right)
\]
and finally $\mathrm{PAPR}\left(
{{\mathchoice{\mbox{\boldmath$\displaystyle s$}}                      {\mbox{\boldmath$\textstyle s$}}                      {\mbox{\boldmath$\scriptstyle s$}}                      {\mbox{\boldmath$\scriptscriptstyle s$}}}%
}\right)  \leq y_{0}$ since $y_{N}\left(  A_{0}^{\ast},\ldots,A_{N-1}^{\ast
}\right)  $ is simply a non-random quantity. Finally, by the LDP $y_{0}%
\leq\log\left(  N\right)  $ for $N$ large enough. Since the expectation are
somewhat difficult to handle instead of the $\mathrm{PAPR}\left(
{{\mathchoice{\mbox{\boldmath$\displaystyle s$}}                      {\mbox{\boldmath$\textstyle s$}}                      {\mbox{\boldmath$\scriptstyle s$}}                      {\mbox{\boldmath$\scriptscriptstyle s$}}}%
}\right)  $ typically the set function and corresponding bounds have been
used. For example, Chernoff bounds have been used in
{\cite{sharif_04_sig,litsyn_06_inf,WangSP08} }showing good performance and low
complexity. Moment bounds with better tail properties have been used but the
complexity is higher {\cite{DamavandiVTC11}}. Performance results of the
derandomization method are reported in Fig. \ref{fig:litsyn} comparing
sequence balancing (Sec. \ref{sec:ldp_part}) with and without derandomization.
The benefit of the derandomization method is clearly observed and corresponds
to more than 4 dB gain in HPA backoff (at $10^{-3}$ outage probability) which
mimics exactly the results of the LDP analysis (Sec. \ref{sec:ldp_ldp}).
However, it comes at the cost of 1 bit/dimension rate loss. The relevant trade
off between rate and PAPR have been rarely investigated so far.

Combining the derandomization method with partitioning yields several improved
algorithms for standard problems. For example the PTS method has been applied
in \cite{sharif_06_toc}. It is proved that with derandomization method PTS can
achieve $r\log\left(  N\right)  $ where $r$ is the percentage of partial
transmit sequences related to $N$. The tone reservation method has been
treated using derandomization in {\cite{litsyn_06_inf}} (see also Sec.
\ref{sec:banach_tr}). Implicitly, derandomization has been used in
\cite{schmidt_10_cl} to show that PAPR of some translate of a code
$\mathcal{C}$ is below $|\mathcal{C}|\log\left(  N\right)  $. Related
derandomization algorithms have been used in \cite{sharif_09_toc} adopting the
so-called pusher-chooser game from \cite{spencer_94}. The idea is to choose
$l_{p}$-norms and prove a recursive formula similar to the Chernoff method.
The approach can be generalized to alternative metrics if appropriate bounds
are available: in \cite{WS10apat} the SER has been reduced using
derandomization showing that $\frac{\log\left(  N\right)  }{2}$ clipping level
is sufficient asymptotically for zero error probability (instead of
$\log\left(  N\right)  $). Furthermore, in the recent paper
\cite{wunder_11_isit} zero clipped energy is asymptotically achieved with
clipping level $\log\log\left(  N\right)  $.

There is still plenty of room for improvements, e.g. by considering
correlations between different samplings points and incorporating other
metrics as well \cite{wunder_11_isit,WS10apat}. It has not been noticed yet
that this field is particularly underdeveloped and bears great potential for
significant improvements of currents systems. Another point to be improved is
the rate loss imposed by the current methods.

\section{Additional resources: MIMO and multiuser systems}

\label{sec:mimo}

While utmost beneficial in terms of spectral efficiency MIMO systems
complicate the PAPR problem: in single-antenna systems the PAPR (or other
metrics) of only one transmit antenna has to be controlled. In the MIMO
setting a large number of OFDM signals are transmitted in parallel and
typically the worst-case candidate dictates the PAPR metric (e.g., due to
out-of-band power) \cite{siegl_2010}.

As a consequence, PAPR reduction methods tailored to this situation should be
utilized, instead of performing single-antenna PAPR reduction in parallel.
Multi-antenna transmitter provide additional degrees of freedom which can be
utilized beneficially for PAPR reduction---the full potential has not yet been
explored in literature. Basically, the peak power can be redistributed over
the antennas. By this, MIMO PAPR reduction may lead to an increased slope of
the CCDF curves (cf.\ Fig~\ref{fig:spm_log}), i.e., the probability of
occurrence of large signal peaks can significantly lowered compared to
single-antenna schemes. This effect is similar to that of achieving some
diversity gain.

When studying MIMO PAPR reduction schemes two basic scenarios have to be
distinguished: on the one hand, in point-to-point MIMO transmission joint
processing of the signals at both ends (transmitter and receiver) is possible.
On the other hand, in point-to-multipoint situations, i.e., multi-user
downlink transmission joint signal processing is only possible at the
transmitter side\footnote{In the \emph{multipoint-to-point scenario}
(multiple-access channel) no joint optimization of the transmit signals can be
performed, hence this case is not amenable for MIMO approaches.}. This fact
heavily restricts the applicability of PAPR reduction schemes.

For the point-to-point setting, a number of PAPR reduction schemes have been
designed, particularly extension of SLM. Besides \emph{ordinary SLM}
(conventional SLM is simply applied in parallel) \emph{simplified SLM} (the
selection is coupled over the antennas) has been proposed in \cite{baek_04}.
\emph{Directed SLM} \cite{fischer_09_toc} is tailored to the MIMO situation
and successively invests complexity (test of candidates) only where PAR
reduction is really needed.

It might be sufficient that the PAPR stays below a tolerable limit, determined
by the actual radio frontend. Here, complexity can be saved if candidate
generation and assessment is done successively and stopped if the tolerable
value is reached. Interestingly, the average number of assessed candidates is
simply given by the inverse of the cdf of PAPR of the underlying original OFDM
scheme. Noteworthy, for $\mathrm{PAPR}=\log(N)$ and reasonably large number
$N$ of carriers, average complexity per antenna is in the order of
$\mathrm{e}=2.718\ldots$ (Euler's number) \cite{siegl_2010sp}. Alternative
metrics have been used in \cite{LIT_ett2010}.

Compared to point-to-point MIMO systems, PAPR reduction schemes applicable in
point-to-multipoint scenarios (multi-user downlink) are a much more
challenging task. Since no joint receiver-side signal processing is possible,
at the transmitter side in candidate generation only operation are allowed
which can individually be reversed at each of the receivers. Among the SLM
family, only simplified SLM can be used here. However, in this situation the
usually present transmitter side multi-user pre-equalization can be utilized
for PAPR reduction. Applying Tomlinson-Harashima precoding the sorting in each
carrier can be optimized to lower PAPR at almost no cost in (uncoded) error
rate \cite{siegl_2010}. The same is true when applying lattice-reduction-aided
pre-equalization. Here the unimodular matrices (describing a change of basis)
can be optimized to control the properties of the transmit signals
\cite{siegl_11}. {There are also links from MIMO PAPR reduction and
derandomization to code design (cf.\ Sec.~\ref{sec:coding}).}

\section{Going beyond: OFDM Capacity Fundamentals}

\begin{figure}[tbh]
\centerline{\includegraphics[width=0.8\columnwidth]{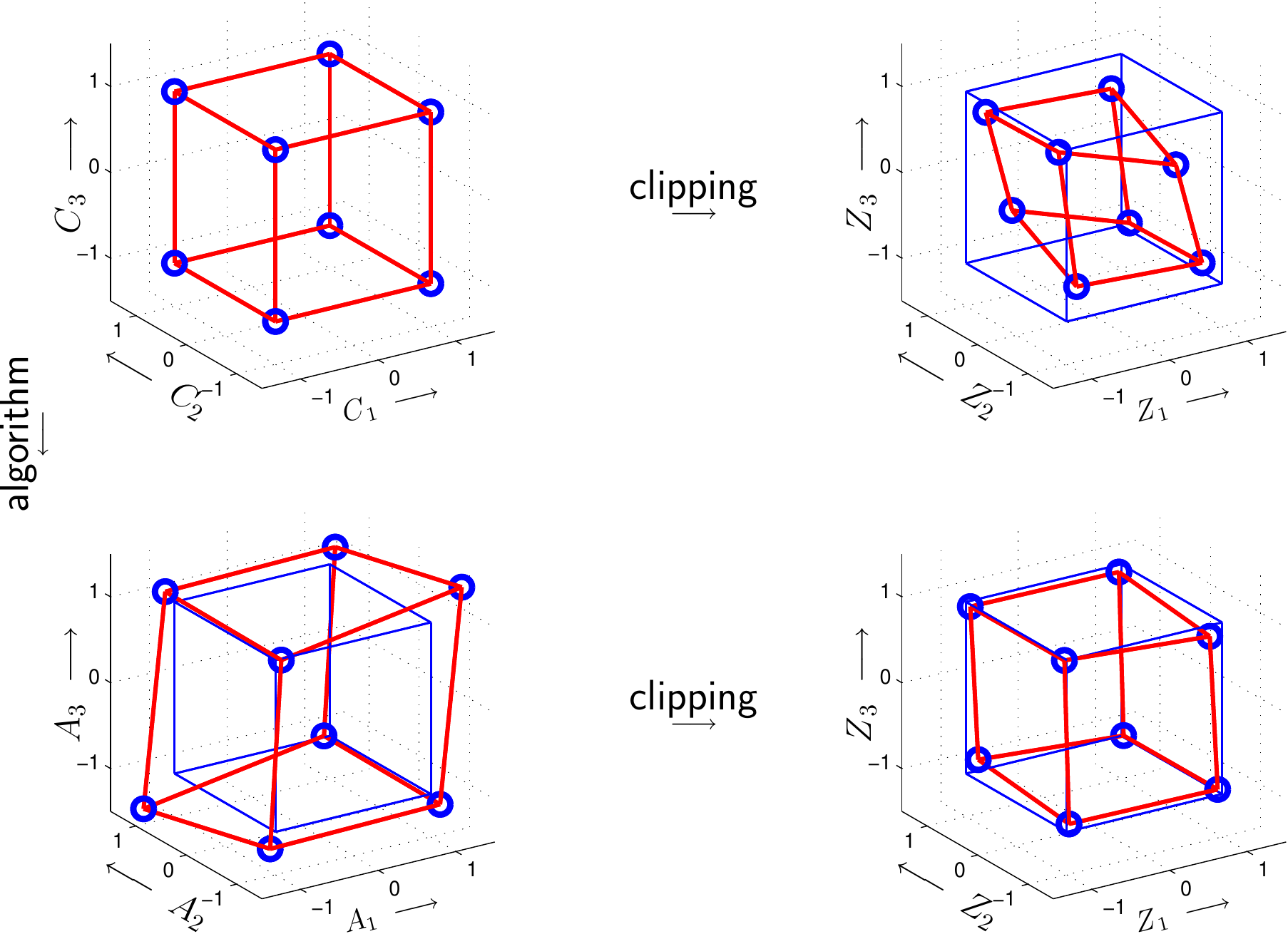}}\caption{Top
row: visualization of the effect of clipping on the set of possible OFDM
frames (here: $N=3$, 2-PAM per carrier). Bottom row: visualization of
predistortion via an algorithm for maximizing the power of the signal after
clipping. }%
\label{fig:clip}%
\end{figure}

While the capacity of the discrete-time peak-power-constraint channel is known
and computable, the capacity of the OFDM peak-power-constraint channel is
still an open problem \cite{smith_71_infc,shamai_95_inf}. The problem is
indeed intricate as it has been unknown until very recently that there are
exponentially many OFDM signals with constant PAPR
{(cf.\ Sec.~\ref{sec:coding})}. However, no practical encoding scheme is known
which comes even close to this merely theoretical result. From this
perspective the capacity problem awaits a more thorough theoretical solution.

{Recent work \cite{peng_06,fischer_10} on practical schemes indicates that
clipped OFDM performs (almost) the same as unclipped OFDM with signal-to-noise
ratio reduced according to the clipping power loss. The main source of the
loss are not the introduced distortions or errors but simply the reduced
output power. Given the OFDM frame in frequency domain
${\mathchoice{\mbox{\boldmath$\displaystyle C$}}                      {\mbox{\boldmath$\textstyle C$}}                      {\mbox{\boldmath$\scriptstyle C$}}                      {\mbox{\boldmath$\scriptscriptstyle C$}}}%
=[C_{1},\ldots,C_{N}]$, via IDFT the time-domain samples $s[k]$ are
calculated. These samples then undergo clipping in the amplifier frontend. As
usually the clipping behavior can be described by a nonlinear, memoryless
point symmetric function $g(x)$ (with $g(x)\leq x$, $x>0$, applied
element-wise to vectors). In frequency domain, the clipped signal is given by
${\mathchoice{\mbox{\boldmath$\displaystyle Z$}}                      {\mbox{\boldmath$\textstyle Z$}}                      {\mbox{\boldmath$\scriptstyle Z$}}                      {\mbox{\boldmath$\scriptscriptstyle Z$}}}%
=\mathrm{DFT}\{g(\mathrm{IDFT}%
\{{\mathchoice{\mbox{\boldmath$\displaystyle C$}}                      {\mbox{\boldmath$\textstyle C$}}                      {\mbox{\boldmath$\scriptstyle C$}}                      {\mbox{\boldmath$\scriptscriptstyle C$}}}%
\})\}$. Note that clipping is a deterministic function and a one-to-one
relation between the vector
${\mathchoice{\mbox{\boldmath$\displaystyle C$}}                      {\mbox{\boldmath$\textstyle C$}}                      {\mbox{\boldmath$\scriptstyle C$}}                      {\mbox{\boldmath$\scriptscriptstyle C$}}}%
$ of unclipped symbols and the vector
${\mathchoice{\mbox{\boldmath$\displaystyle Z$}}                      {\mbox{\boldmath$\textstyle Z$}}                      {\mbox{\boldmath$\scriptstyle Z$}}                      {\mbox{\boldmath$\scriptscriptstyle Z$}}}%
=[Z_{1},\ldots,Z_{N}]$ of clipped ones exist. Assuming an AWGN channel, at the
receiver side the vector
${\mathchoice{\mbox{\boldmath$\displaystyle Z$}}                      {\mbox{\boldmath$\textstyle Z$}}                      {\mbox{\boldmath$\scriptstyle Z$}}                      {\mbox{\boldmath$\scriptscriptstyle Z$}}}%
$, disturbed by additive white Gaussian noise is present. In case of
intersymbol-interference channels, the symbols $Z_{n}$ are additionally
individually scaled by the fading gain at the respective carrier. }

This clipping behavior can be visualized for $N=3$ and 2-PAM per carrier, see
top of Fig.~\ref{fig:clip}. The initial hypercube with vertices given by all
possible vectors
${\mathchoice{\mbox{\boldmath$\displaystyle C$}}                      {\mbox{\boldmath$\textstyle C$}}                      {\mbox{\boldmath$\scriptstyle C$}}                      {\mbox{\boldmath$\scriptscriptstyle C$}}}%
$ is distorted. However, the attenuation of the useful signal (the vector
${\mathchoice{\mbox{\boldmath$\displaystyle Z$}}                      {\mbox{\boldmath$\textstyle Z$}}                      {\mbox{\boldmath$\scriptstyle Z$}}                      {\mbox{\boldmath$\scriptscriptstyle Z$}}}%
$ has lower energy) will be the dominating effect over deformation. This, in
turn, leads to the conclusion that a suited metric for capacity maximization
is simply the average power of the power amplifier output signal.

A possible strategy is shown on the bottom of Fig.~\ref{fig:clip}. A signal
shaping algorithm may adjust the signal points in $2N$-dimensional real-valued
space such that after clipping the set of all possible OFDM vectors in
frequency domain forms (approximately) an hypercube with energy close to that
of the initial constellation. First work on using the strategy of \emph{active
constellation extension} for achieving is goal has been presented
\cite{siegl_inowo10}.

\section{Emerging solutions: An open field}

\subsection{New Trends in Code Design}

\label{sec:coding}

Jones et al.\ \cite{jones_94_el} were the first to describe block coding
schemes in the present context. This framework has been put in systematic form
by observing the connection of cosets of Reed-Muller codes and complementary
sequences \cite{davis_99_inf,schmidt_06_inf}. Unfortunately, these approaches
have limited potential for modern OFDM systems due to their limited coding
rate. The fundamental trade off between different code key properties such as
rate, PAPR etc. was explored and discussed in \cite{paterson_00_inf2}. More
recent ideas use the idea of sequence balancing and code extensions in form of
erasure coding in other domains (e.g.\ MIMO \cite{fischer_09_inf}) to tackle
the PAPR problem with an inner code, while error correction still is done via
an outer code \cite{litsyn_07_toc}.

\subsubsection{Codes and sequences with low PAPR}

Though most of multicarrier signals of length $N$ have PAPR close to
$\log\left(  N\right)  $, it turns out that signal with constant PAPR are not
so rare. Using a remarkable result of Spencer \cite{spen85} it is possible to
show that the number of such BPSK modulated signals is exponential in $N$.
Namely, there are at least $(2-\delta_{K})^{N}$ such signals with PAPR not
exceeding $K$, where $\delta_{K}$ is a constant depending on $K$ and tending
to zero when $K$ grows. It is an open question how to generate many signals
for given $K$.

A lot of research was devoted to describing signals with low values of PAPR.
For BPSK modulated signals an extreme example is provided by Rudin-Shapiro
sequences defined recursively from $P_{0}=Q_{0}=1$, and
\[
P_{m+1}=(P_{m},Q_{m}),\ \ Q_{m+1}=(P_{m},-Q_{m}).
\]
These sequences of length being a power of $2$ have PAPR at most $2$. More
general examples of sequences with PAPR at most 2 arise from Golay
complementary sequences. Two sequences constitute a complementary pair if the
sum of the values of their aperiodic correlation functions sum up to zero.
Many methods are known for constructing such sequences, see \cite[Section
7.6]{litsyn_07}. Notice that it is not known if BPSK modulated signals can
have PAPR less than 2. However, if one increases the size of multiphase
constellations to infinity there exist sequences with PAPR approaching 1
\cite[Theorem 7.37]{litsyn_07}. For constructions of multiphase complementary
pairs from cosets of Reed-Muller codes see \cite{davi99} and references there.
PAPR of $m$-sequences and Legendre sequences is discussed in \cite[Sections
7.7 and 7.8]{litsyn_07}.

Often we need to know the biggest PAPR among sequences belonging to a code.
Bounds on PAPR of codes on sphere as a function of their sizes and minimum
Euclidean distances was studied in \cite{paterson_00_inf2}. A relation between
the distance distribution of codes and PAPR was derived in
\cite{litsyn_06_inf}. This yielded bounds on PAPR of long algebraic codes,
such as BCH codes. Analysis of PAPR of codes with iterative decoding, for
instance LDPC codes, remains an open problem. PAPR of codes of small size was
studied in \cite{paterson_00_inf2}. In particular, it was shown that PAPR of
duals of length $N$ BCH codes are at most $\log^{2}\left(  N\right)  $. As
well bounds on PAPR of Kerdock and Delsarte-Goethals codes were derived. In
\cite{litsyn_07_toc} it was shown that in every coset of a code dual to BCH
code with the minimum distance of $\log\left(  N\right)  $ exists a code word
with PAPR at most $\log\left(  N\right)  $. At the same time, this leads to a
very modest rate loss. Still, constructing codes having low PAPR and high
minimum distance seems to be a challenge.

Computing PAPR of a given code is a computationally consuming problem. If a
code has a reasonably simple maximum-likelihood decoding algorithm it is
possible to determine efficiently its PAPR \cite{wunder_02_itg,taro00}.

\begin{figure}[t]
\centering
\scalebox{.9}{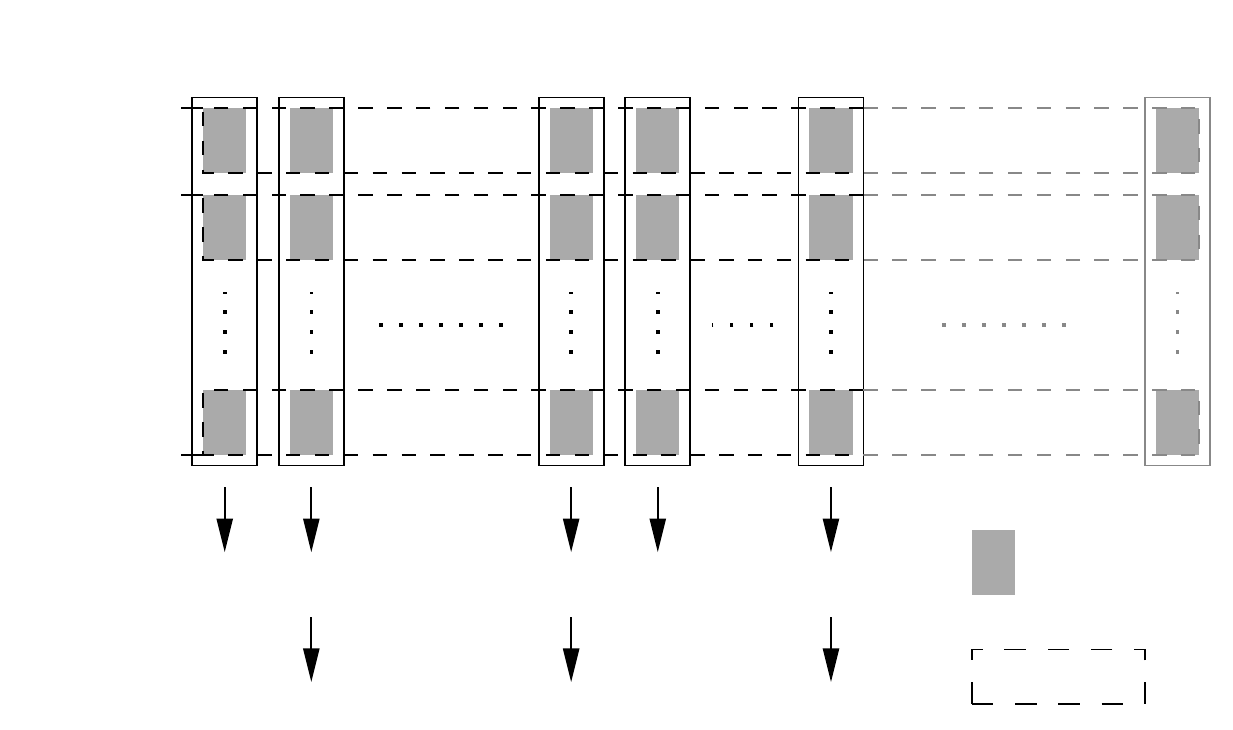}\caption{Illustration of the
coding method across symbols using off-the-shelf codes (such as RS codes).}%
\label{fig:simplex}%
\end{figure}

{In \cite{fischer_09_inf} off-the-shelf channel codes, in particular
Reed--Solomon (RS) and Simplex codes are employed to create candidates, from
which, as in SLM, the best are selected. The codes are thereby arranged over a
number of OFDM frames rather than over the carriers. Such an approach is very
flexible as due to the selection step any criterion of optimality can be taken
into account. Moreover, instead of applying the approach to the MIMO setting,
it can also be used if block of temporal consecutive OFDM frames are treated
jointly. The method is illustrated in Fig. \ref{fig:simplex}.}

\subsubsection{Constellation shaping}

In constellation shaping, we have to find a constellation in the
$N$-dimensional frequency domain, such that the resulting shaping region in
the time domain has low PAPR. At the same time we would like to have a simple
encoding method for the chosen constellation. Such shaping based on Hadamard
transform was considered in \cite{kwok00}. The main challenge in constellation
shaping is to find a unique way of mapping (encoding) and its inverse
(decoding) of reasonable complexity. The suggested approach in
\cite{mobasher06,kwok00} is based on a matrix decomposition. Though the
simulation results are quite\ promising, the implementation complexity still
seems to be far from being affordable \cite{eugen10,mobasher06}.

\subsection{Banach space geometry}

\label{sec:banach}

An interesting new approach to the PAPR problem is that of using Banach space
geometry. Banach space geometry relates norms and metrics of different Banach
spaces to each other. For example, a question that often arises is: assume a
Banach space with unit norm ball $B_{1}$ and another Banach space with ball
$B_{2}$; both spaces are of finite, possibly different dimension. What is the
relation between the norms if the projection of one ball covers the other
ball? Furthermore, what is the dependence of this relation on the dimensions?

Interestingly, these relations turn out to be useful for the PAPR problem in
several other ways depending on the underlying Banach spaces as the following
examples show:

\subsubsection{Alternative orthonormal systems: Kashin \& Tzafriri's theorem}

In Sec. \ref{sec:ldp} it was shown that OFDM has unfavorable PAPR of order
$\log\left(  N\right)  $ if $N$ gets large. One might be inclined to ask if
this is an artifact of the underlying orthonormal signaling system. The answer
is actually no with the implication that OFDM plays no specific role among all
orthonormal systems. Already in \cite{pohl_07_ett} it was shown that worst
PAPR is of order $N$ regardless of the signaling system (multicarrier CDM
etc.). But even if we consider not the worst PAPR but look at the PAPR on
average the situation does not get better. In \cite{kashin95}, Kashin \&
Tzafriri proved that for any orthonormal system on a given finite time
interval the expectation of PAPR is necessarily of order $\log\left(
N\right)  $. Again, changing the signaling is not beneficial in terms of PAPR.
The underlying mathematical problem is that of estimating the supremum norm of
a finite linear combination of functions weighted with random coefficients
both constrained in the energy norm.

\subsubsection{Is PAPR of single-carrier really much better?}

It is common engineering experience that single-carrier has better PAPR than
multicarrier. But it might be worth raising this question again within the
context of upcoming technological advances (LTE-A etc.) which operate much
closer to the Nyquist bandwidth and, moreover, use different modulation and
coding schemes. Let us formalize this question.

Suppose, we send a transmit sequence $C_{1},...,C_{N}$ and use a band-limited
filter to generate the continuous-time signal (bandwidth is set to $\pi$ for
simplicity). The transmit signal can be described by
\[
s\left(  t\right)  =\sum_{i=1}^{N}C_{i}\frac{\sin\pi\left(  t-t_{i}\right)
}{\pi\left(  t-t_{i}\right)  }%
\]
with sampling points $t_{i}\in\mathbb{Z}$. Naturally, band-limited signals of
this form have very different PAPR behavior compared to OFDM since, obviously,
if the coefficients are from some standard modulation alphabet, the signal is
nailed down to some finite value at the sampling point independent of $N$.
However, within the sampling intervals (on average) large PAPR could actually
occur. Noteworthy, the worst case is growing without bounds linearly in $N$.

Surprisingly, the exact answer to this problem has not been explored until
very recently \cite{Boche2011_JFAA} which is basically a result on large
deviations in Banach spaces. It is proved in \cite{Boche2011_JFAA} that such
bad PAPR cannot actually happen and that there is a constant $c_{0}>0$ such
that:%
\[
\mathbb{E}\left(  \mathrm{PAPR}\right)  \leq c_{0}\log\log\left(  N\right)
\]
But we also see the catch here. Modern communication systems use higher
modulation sizes and in that case the influence of the data becomes dominant
if the distribution becomes Gaussian like. In that case we approach the
log$\left(  N\right)  $ again.

There is some interesting connection of the PAPR problem to the Hilbert
transform context: since in many standard communication models, e.g. in
Gabor's famous \emph{Theory of Communication} \cite{gabor46,bm12_sse}, the
transmit signal is a linear combination of a signal and its Hilbert transform,
properties such as PAPR in the transform domain become more and more
important. Initiated by early works of Logan \cite{logan78} who investigated
the Hilbert transforms of certain bandpass signals it was recogniced not until
very recently \cite{bm12_inft,bm12_icassp} that the results are fragile for
wideband signals containing spectral components in an interval around zero
frequency. Then, in general, the domain of the Hilbert transform must be
suitably extended; further, examples of bandlimited wideband signals are
provided where the PAPR grows without bounds in the Hilbert transform domain
\cite{bm12_eusipco}. Hence, for certain single-carrier analytic modulation
schemes the transmit signal has to be shaped very carefully.

\subsubsection{Overcomplete expansions with uniformly bounded PAPR}

While the result for arbitrary orthonormal systems appears rather pessimistic
there is a possible solution in the form of \emph{frames}. Frames are
overcomplete systems of vectors in $\mathbb{R}^{n},n<N$. Let us denote this
description by
{${\mathchoice{\mbox{\boldmath$\displaystyle U$}}                      {\mbox{\boldmath$\textstyle U$}}                      {\mbox{\boldmath$\scriptstyle U$}}                      {\mbox{\boldmath$\scriptscriptstyle U$}}}%
$}%
$=[{{\mathchoice{\mbox{\boldmath$\displaystyle u$}}                      {\mbox{\boldmath$\textstyle u$}}                      {\mbox{\boldmath$\scriptstyle u$}}                      {\mbox{\boldmath$\scriptscriptstyle u$}}}%
}_{1}%
,...,{{\mathchoice{\mbox{\boldmath$\displaystyle u$}}                      {\mbox{\boldmath$\textstyle u$}}                      {\mbox{\boldmath$\scriptstyle u$}}                      {\mbox{\boldmath$\scriptscriptstyle u$}}]}%
}_{N}\in\mathbb{R}^{n\times N},N\geq n$. Then, if the rows are independent
there is
{${\mathchoice{\mbox{\boldmath$\displaystyle x$}}                      {\mbox{\boldmath$\textstyle x$}}                      {\mbox{\boldmath$\scriptstyle x$}}                      {\mbox{\boldmath$\scriptscriptstyle x$}}}%
$}$\in\mathbb{R}^{N}$ so that%
\begin{equation}
{{\mathchoice{\mbox{\boldmath$\displaystyle y$}} {\mbox{\boldmath$\textstyle y$}} {\mbox{\boldmath$\scriptstyle y$}} {\mbox{\boldmath$\scriptscriptstyle y$}}}%
}%
={{\mathchoice{\mbox{\boldmath$\displaystyle U$}} {\mbox{\boldmath$\textstyle U$}} {\mbox{\boldmath$\scriptstyle U$}} {\mbox{\boldmath$\scriptscriptstyle U$}}}%
}^{T}%
{{\mathchoice{\mbox{\boldmath$\displaystyle x$}} {\mbox{\boldmath$\textstyle x$}} {\mbox{\boldmath$\scriptstyle x$}} {\mbox{\boldmath$\scriptscriptstyle x$}}}%
} \label{eqn:U_map}%
\end{equation}
for any
${{\mathchoice{\mbox{\boldmath$\displaystyle y$}}                      {\mbox{\boldmath$\textstyle y$}}                      {\mbox{\boldmath$\scriptstyle y$}}                      {\mbox{\boldmath$\scriptscriptstyle y$}}}%
}=\mathbb{R}^{n}$ and the elements of
{${\mathchoice{\mbox{\boldmath$\displaystyle U$}}                      {\mbox{\boldmath$\textstyle U$}}                      {\mbox{\boldmath$\scriptstyle U$}}                      {\mbox{\boldmath$\scriptscriptstyle U$}}}%
$} are a frame. If
{${\mathchoice{\mbox{\boldmath$\displaystyle U$}}                      {\mbox{\boldmath$\textstyle U$}}                      {\mbox{\boldmath$\scriptstyle U$}}                      {\mbox{\boldmath$\scriptscriptstyle U$}}}%
$}$^{T}%
{{\mathchoice{\mbox{\boldmath$\displaystyle U$}}                      {\mbox{\boldmath$\textstyle U$}}                      {\mbox{\boldmath$\scriptstyle U$}}                      {\mbox{\boldmath$\scriptscriptstyle U$}}}%
}%
={{\mathchoice{\mbox{\boldmath$\displaystyle I$}}                      {\mbox{\boldmath$\textstyle I$}}                      {\mbox{\boldmath$\scriptstyle I$}}                      {\mbox{\boldmath$\scriptscriptstyle I$}}}%
}_{n}$, where
${{\mathchoice{\mbox{\boldmath$\displaystyle I$}}                      {\mbox{\boldmath$\textstyle I$}}                      {\mbox{\boldmath$\scriptstyle I$}}                      {\mbox{\boldmath$\scriptscriptstyle I$}}}%
}_{n}$ is the identity matrix, then it is called a \emph{tight frame}. In
seminal work Kashin \cite{Kashin} interpreted the mapping (\ref{eqn:U_map}) as
an embedding of the Banach space with supremum norm $l_{\infty}$ to the Banach
space with standard euclidean norm $l_{2}$ and asked for the growth factor
$K\left(  \lambda\right)  >0,\lambda:=N/n$, between the two norms when the
$l_{2}$ unit ball in $\mathbb{R}^{n}$ should be covered by the unit ball
$l_{\infty}$ in $\mathbb{R}^{N}$. Such representations are called \emph{Kashin
representations} of level $\lambda$ \cite{farrel_09}.

Clearly, if $N=n$ then $K\left(  \lambda\right)  =\sqrt{N}$. However, if $N>n$
(overcomplete expansion) then Kashin proved that there is a subspace in
$\mathbb{R}^{N}$ generating by a frame
${{\mathchoice{\mbox{\boldmath$\displaystyle U$}}                      {\mbox{\boldmath$\textstyle U$}}                      {\mbox{\boldmath$\scriptstyle U$}}                      {\mbox{\boldmath$\scriptscriptstyle U$}}}%
}$ such that $\lambda$ is given by:%
\[
K\left(  \lambda\right)  :=c_{1}\left(  \frac{\lambda}{\lambda-1}\log\left(
1+\frac{\lambda}{\lambda-1}\right)  \right)  ^{1/2},\;c_{1}>0
\]
Hence, the $K\left(  \lambda\right)  $ is uniformly bounded in $n$ if
$\lambda>1$ is fixed. Good estimates of the constant $c_{1}>0$ are not known
\cite{farrel_09}.

This intriguing result has been applied in the PAPR context in
\cite{farrel_09} and the implications for peak power reduction are immediate.
The matrix
${{\mathchoice{\mbox{\boldmath$\displaystyle U$}}                      {\mbox{\boldmath$\textstyle U$}}                      {\mbox{\boldmath$\scriptstyle U$}}                      {\mbox{\boldmath$\scriptscriptstyle U$}}}%
}$ can be taken as a precoding matrix for classical OFDM transmission and
achieve uniformly bounded PAPR. Unfortunately the construction of the optimal
subspace is not known \cite{farrel_09}. Kashin representations exploiting the
\emph{uncertainity principle} of random partial Fourier matrices are presented
in \cite{Ilic09}.

\subsubsection{Tone reservation and Szemer\'{e}di's theorem}

\label{sec:banach_tr}

One of the oldest but still very popular scheme is tone reservation
\cite{wang12}. But, despite its simplicity, many questions involved are still
open which does not come by coincidence: recent work in
\cite{Boche2011_EURASIP} has analyzed the performance of this method and uses
an application of the Szemer\'{e}di's famous theorem about arithmetic
progressions (Abel price 2012).

Recalling the setting where a subset of subcarriers is solely reserved for
peak power reduction the challenge is to find for a given set of transmit
sequence a subset and corresponding values such that the PAPR is reduced to
the most possible gain. Until now, achievability and limits are not known
(except for simple cases). Therefore, there is some incentive to look at this
problem from a new perspective. Ref. \cite{Boche2011_EURASIP} has analyzed the
case where the compensation set is arbitrary but fixed. In this typical case
it is proved that the efficiency of the system, i.e., the ratio of cardinality
of information and compensation sets must decrease to zero if the peak power
is constrained independent of the subcarriers. The technique that is used is
to show necessary assumptions on the relations of unit spheres in the Banach
spaces. This relation is shown not to hold asymptotically for sets with
additive structure. However, Szemer\'{e}di's theorem states that such sets are
included in every subset of cardinality $\delta N$ where $\delta>0$. In fact
such arithmetic progressions induce signals with bad PAPR behavior naturally
to be excluded by the method. The theorem shows that this is not possible.

In extended work \cite{boche12} also other families of orthogonal signalling
such as Walsh sequences are analyzed all of them showing basically the same
disencouraging result regarding the system's efficiency. This leads to the
conjecture in \cite{boche12} that all natural orthogonal signalling families
have this behavior.

\subsection{Compressed sensing}

Compressed sensing \cite{Donoho1,Candes1} is a new sampling method that
compresses a signal simultaneously with data acquisition. Each element of the
compressed signal or measurements consists of a linear combination of the
elements in the original signal and this linear transformation is independent
of instantaneous characteristics of each signal. In general, it is not
possible to recover an unknown original signal from the measurements in the
reduced dimension. Nevertheless, if the original signal has sparsity property,
its recovery can be perfectly achieved at the receiver. Since sparsity
frequently appears in the PAPR problems of the OFDM systems, compressed
sensing can be a powerful tool to solve these problems.

Compressed sensing can be regarded as minimizing the number of measurements
while still retaining the information necessary to recover the original signal
well (i.e. beyond classical Nyquist sampling). The process can be briefly
illustrated as follows. Let
${\mathchoice{\mbox{\boldmath$\displaystyle f$}}                      {\mbox{\boldmath$\textstyle f$}}                      {\mbox{\boldmath$\scriptstyle f$}}                      {\mbox{\boldmath$\scriptscriptstyle f$}}}%
$ denote a signal vector of dimension $N$ and
${\mathchoice{\mbox{\boldmath$\displaystyle g$}}                      {\mbox{\boldmath$\textstyle g$}}                      {\mbox{\boldmath$\scriptstyle g$}}                      {\mbox{\boldmath$\scriptscriptstyle g$}}}%
$ be a measurement vector of dimension $M$ with $M<N$ obtained by
${\mathchoice{\mbox{\boldmath$\displaystyle g$}}                      {\mbox{\boldmath$\textstyle g$}}                      {\mbox{\boldmath$\scriptstyle g$}}                      {\mbox{\boldmath$\scriptscriptstyle g$}}}%
={\mathchoice{\mbox{\boldmath$\displaystyle \Phi$}}                      {\mbox{\boldmath$\textstyle \Phi$}}                      {\mbox{\boldmath$\scriptstyle \Phi$}}                      {\mbox{\boldmath$\scriptscriptstyle \Phi$}}}%
{\mathchoice{\mbox{\boldmath$\displaystyle f$}}                      {\mbox{\boldmath$\textstyle f$}}                      {\mbox{\boldmath$\scriptstyle f$}}                      {\mbox{\boldmath$\scriptscriptstyle f$}}}%
$, where
${\mathchoice{\mbox{\boldmath$\displaystyle \Phi$}}                      {\mbox{\boldmath$\textstyle \Phi$}}                      {\mbox{\boldmath$\scriptstyle \Phi$}}                      {\mbox{\boldmath$\scriptscriptstyle \Phi$}}}%
$ is called \emph{sensing matrix}. At the transmitter, sampling and
compression are performed altogether by simply multiplying
${\mathchoice{\mbox{\boldmath$\displaystyle \Phi$}}                      {\mbox{\boldmath$\textstyle \Phi$}}                      {\mbox{\boldmath$\scriptstyle \Phi$}}                      {\mbox{\boldmath$\scriptscriptstyle \Phi$}}}%
$ by
${\mathchoice{\mbox{\boldmath$\displaystyle f$}}                      {\mbox{\boldmath$\textstyle f$}}                      {\mbox{\boldmath$\scriptstyle f$}}                      {\mbox{\boldmath$\scriptscriptstyle f$}}}%
$ to obtain
${\mathchoice{\mbox{\boldmath$\displaystyle g$}}                      {\mbox{\boldmath$\textstyle g$}}                      {\mbox{\boldmath$\scriptstyle g$}}                      {\mbox{\boldmath$\scriptscriptstyle g$}}}%
$. At the receiver, if
${\mathchoice{\mbox{\boldmath$\displaystyle f$}}                      {\mbox{\boldmath$\textstyle f$}}                      {\mbox{\boldmath$\scriptstyle f$}}                      {\mbox{\boldmath$\scriptscriptstyle f$}}}%
$ is an $S$-sparse signal, which means
${\mathchoice{\mbox{\boldmath$\displaystyle f$}}                      {\mbox{\boldmath$\textstyle f$}}                      {\mbox{\boldmath$\scriptstyle f$}}                      {\mbox{\boldmath$\scriptscriptstyle f$}}}%
$ has no more than $S$ nonzero elements, it is shown in \cite{Candes1} that
the exact
${\mathchoice{\mbox{\boldmath$\displaystyle f$}}                      {\mbox{\boldmath$\textstyle f$}}                      {\mbox{\boldmath$\scriptstyle f$}}                      {\mbox{\boldmath$\scriptscriptstyle f$}}}%
$ can be obtained from
${\mathchoice{\mbox{\boldmath$\displaystyle g$}}                      {\mbox{\boldmath$\textstyle g$}}                      {\mbox{\boldmath$\scriptstyle g$}}                      {\mbox{\boldmath$\scriptscriptstyle g$}}}%
$ by using $l_{1}$ minimization, that is,%
\begin{equation}
\min
_{{\mathchoice{\mbox{\boldmath$\displaystyle \tilde{f}$}} {\mbox{\boldmath$\textstyle \tilde{f}$}} {\mbox{\boldmath$\scriptstyle \tilde{f}$}} {\mbox{\boldmath$\scriptscriptstyle \tilde{f}$}}}%
}%
{||{\mathchoice{\mbox{\boldmath$\displaystyle \tilde{f}$}} {\mbox{\boldmath$\textstyle \tilde{f}$}} {\mbox{\boldmath$\scriptstyle \tilde{f}$}} {\mbox{\boldmath$\scriptscriptstyle \tilde{f}$}}}%
||_{1}}\quad\mathrm{subject~to}\quad
{\mathchoice{\mbox{\boldmath$\displaystyle g$}} {\mbox{\boldmath$\textstyle g$}} {\mbox{\boldmath$\scriptstyle g$}} {\mbox{\boldmath$\scriptscriptstyle g$}}}%
={\mathchoice{\mbox{\boldmath$\displaystyle \Phi$}} {\mbox{\boldmath$\textstyle \Phi$}} {\mbox{\boldmath$\scriptstyle \Phi$}} {\mbox{\boldmath$\scriptscriptstyle \Phi$}}}%
{\mathchoice{\mbox{\boldmath$\displaystyle \tilde{f}$}} {\mbox{\boldmath$\textstyle \tilde{f}$}} {\mbox{\boldmath$\scriptstyle \tilde{f}$}} {\mbox{\boldmath$\scriptscriptstyle \tilde{f}$}}}
\label{l1_min}%
\end{equation}
as long as
${\mathchoice{\mbox{\boldmath$\displaystyle \Phi$}}                      {\mbox{\boldmath$\textstyle \Phi$}}                      {\mbox{\boldmath$\scriptstyle \Phi$}}                      {\mbox{\boldmath$\scriptscriptstyle \Phi$}}}%
$ has some good property, which is called \emph{restricted isometry property
(RIP)}. For some positive integer $S$, the isometry constant $\delta_{S}$ of a
matrix
${\mathchoice{\mbox{\boldmath$\displaystyle \Phi$}}                      {\mbox{\boldmath$\textstyle \Phi$}}                      {\mbox{\boldmath$\scriptstyle \Phi$}}                      {\mbox{\boldmath$\scriptscriptstyle \Phi$}}}%
$ is defined as the smallest number such that
\[
(1-\delta_{S}%
)||{\mathchoice{\mbox{\boldmath$\displaystyle f$}} {\mbox{\boldmath$\textstyle f$}} {\mbox{\boldmath$\scriptstyle f$}} {\mbox{\boldmath$\scriptscriptstyle f$}}}%
||_{2}^{2}\leq
||{\mathchoice{\mbox{\boldmath$\displaystyle \Phi$}} {\mbox{\boldmath$\textstyle \Phi$}} {\mbox{\boldmath$\scriptstyle \Phi$}} {\mbox{\boldmath$\scriptscriptstyle \Phi$}}}%
{\mathchoice{\mbox{\boldmath$\displaystyle f$}} {\mbox{\boldmath$\textstyle f$}} {\mbox{\boldmath$\scriptstyle f$}} {\mbox{\boldmath$\scriptscriptstyle f$}}}%
||_{2}^{2}\leq(1+\delta_{S}%
)||{\mathchoice{\mbox{\boldmath$\displaystyle f$}} {\mbox{\boldmath$\textstyle f$}} {\mbox{\boldmath$\scriptstyle f$}} {\mbox{\boldmath$\scriptscriptstyle f$}}}%
||_{2}^{2}%
\]
holds for all $S$-sparse vectors
${\mathchoice{\mbox{\boldmath$\displaystyle f$}}                      {\mbox{\boldmath$\textstyle f$}}                      {\mbox{\boldmath$\scriptstyle f$}}                      {\mbox{\boldmath$\scriptscriptstyle f$}}}%
$. Under RIP with $\delta_{2S}<\sqrt{2}-1$, (\ref{l1_min}) gives the exact
solution for
${\mathchoice{\mbox{\boldmath$\displaystyle f$}}                      {\mbox{\boldmath$\textstyle f$}}                      {\mbox{\boldmath$\scriptstyle f$}}                      {\mbox{\boldmath$\scriptscriptstyle f$}}}%
$ \cite{Candes2}. This recovery method using $l_{1}$-minimization is called
basis pursuit (BP) \cite{Chen}, which requires high computational complexity.
Many greedy algorithms \cite{Tropp,Needell1,Needell2,Dai} have been developed
to reduce the recovery complexity.

In many applications of compressed sensing such as communication systems, it
is required to recover
${\mathchoice{\mbox{\boldmath$\displaystyle f$}}                      {\mbox{\boldmath$\textstyle f$}}                      {\mbox{\boldmath$\scriptstyle f$}}                      {\mbox{\boldmath$\scriptscriptstyle f$}}}%
$ from the corrupted measurements
${\mathchoice{\mbox{\boldmath$\displaystyle g$}}                      {\mbox{\boldmath$\textstyle g$}}                      {\mbox{\boldmath$\scriptstyle g$}}                      {\mbox{\boldmath$\scriptscriptstyle g$}}}%
^{\prime}%
={\mathchoice{\mbox{\boldmath$\displaystyle g$}}                      {\mbox{\boldmath$\textstyle g$}}                      {\mbox{\boldmath$\scriptstyle g$}}                      {\mbox{\boldmath$\scriptscriptstyle g$}}}%
+{\mathchoice{\mbox{\boldmath$\displaystyle z$}}                      {\mbox{\boldmath$\textstyle z$}}                      {\mbox{\boldmath$\scriptstyle z$}}                      {\mbox{\boldmath$\scriptscriptstyle z$}}}%
$, where
${\mathchoice{\mbox{\boldmath$\displaystyle z$}}                      {\mbox{\boldmath$\textstyle z$}}                      {\mbox{\boldmath$\scriptstyle z$}}                      {\mbox{\boldmath$\scriptscriptstyle z$}}}%
$ is a noise vector of dimension $M$. For this, recovery algorithms such as
basis pursuit denoising \cite{Chen}, Lasso \cite{Tibshirani}, and their
variants have been developed while the existing recovery algorithms can also
be used. However, these algorithms do not still show good performance enough
to be adopted in wireless communication systems which usually require very low
error rate even in severely noisy environments.

Related to PAPR problems, the properties lying in the compressed sensing such
as sparsity, RIP, and recovery algorithms can be utilized in many PAPR
reduction schemes. In \cite{Ilic1} and \cite{Ilic2}, a new tone-reservation
scheme is proposed, which is different from the existing tone-reservation
\cite{tellado_99_phd} in that it provides a guaranteed upper bound for PAPR
reduction as well as guaranteed rates of convergence. This scheme exploits the
RIP of the partial DFT matrix. In \cite{Ding}, a novel convex optimization
approach is proposed to numerically determine the near-optimal tone-injection
solution. Generally, tone-injection \cite{tellado_99_phd} is an effective
approach to mitigate PAPR problem without incurring bandwidth loss. However,
due to its computational complexity, finding the optimal tone-injection
becomes intractable for OFDM systems with a large number of subcarriers.
Therefore, a semi-definite relaxation needs to be adopted in the convex
optimization \cite{Boyd}. Moreover, based on the observation that only a small
number of subcarrier symbols are usually moved, $l_{0}$ minimization is
required and naturally it can be relaxed to $l_{1}$ minimization similar to
compressed sensing literature.

One of the popular solutions to PAPR reduction is clipping the amplitude of
the OFDM signal although the clipping increases the noise level by inducing a
clipping noise. Due to the sparsity of the clipping noise, compressed sensing
can be used to recover and cancel the clipping noise. Before the clipping
noise cancellation schemes using compressed sensing appear, some foundations
of them have been presented. An impulse noise cancellation system using sparse
recovery is firstly proposed in \cite{INbyCS}. In practical systems, there
exists a set of null tones not used for information transmission, which is
exploited as measurements to estimate the impulse noise in time domain at the
receiver. As an extension to \cite{INbyCS}, an alternative recovery algorithm
with low complexity is proposed in \cite{OC}, which exploits the structure of
DFT matrix and available a-priori information jointly for sparse signal
recovery. In \cite{Lampe}, the work in \cite{INbyCS} is extended to the case
of bursty impulse noise whose recovery is based on the application of
block-based compressed sensing. Secondly, a clipping noise cancellation scheme
using frame theory is proposed in \cite{Valbonesi}. Although this scheme uses
not compressed sensing but frame expansion, the frame expansion can be viewed
as a special case of compressed sensing problem with known positions of
nonzero elements. Some additional reserved tones not including data are padded
and they are used as the measurements to recover the clipping noise at the receiver.

Motivated by the above works, clipping noise cancellation schemes using
compressed sensing have been proposed in \cite{Al-Safadi} and its extended
version in \cite{Al-Safadi2}. In \cite{Al-Safadi,Al-Safadi2}, $M$ reserved
tones are allocated before clipping at the transmitter and they cause some
data rate loss. These reserved tones can be exploited as measurements instead
of null tones in {\cite{INbyCS,OC,Lampe,Lampe_2011}}. Let us denote the
transceiver model in frequency domain with clipping noise as
\begin{equation}
{\mathchoice{\mbox{\boldmath$\displaystyle Y$}} {\mbox{\boldmath$\textstyle Y$}} {\mbox{\boldmath$\scriptstyle Y$}} {\mbox{\boldmath$\scriptscriptstyle Y$}}}%
={\mathchoice{\mbox{\boldmath$\displaystyle H$}} {\mbox{\boldmath$\textstyle H$}} {\mbox{\boldmath$\scriptstyle H$}} {\mbox{\boldmath$\scriptscriptstyle H$}}}%
\left(
{\mathchoice{\mbox{\boldmath$\displaystyle C$}} {\mbox{\boldmath$\textstyle C$}} {\mbox{\boldmath$\scriptstyle C$}} {\mbox{\boldmath$\scriptscriptstyle C$}}}%
+{\mathchoice{\mbox{\boldmath$\displaystyle D$}} {\mbox{\boldmath$\textstyle D$}} {\mbox{\boldmath$\scriptstyle D$}} {\mbox{\boldmath$\scriptscriptstyle D$}}}%
\right)
+{\mathchoice{\mbox{\boldmath$\displaystyle Z$}} {\mbox{\boldmath$\textstyle Z$}} {\mbox{\boldmath$\scriptstyle Z$}} {\mbox{\boldmath$\scriptscriptstyle Z$}}}
\label{Alsa1}%
\end{equation}
where
${\mathchoice{\mbox{\boldmath$\displaystyle C$}} {\mbox{\boldmath$\textstyle C$}} {\mbox{\boldmath$\scriptstyle C$}} {\mbox{\boldmath$\scriptscriptstyle C$}}}%
$ and
${\mathchoice{\mbox{\boldmath$\displaystyle Y$}}                      {\mbox{\boldmath$\textstyle Y$}}                      {\mbox{\boldmath$\scriptstyle Y$}}                      {\mbox{\boldmath$\scriptscriptstyle Y$}}}%
$ are $N\times1$ transmitted and received tone vectors, respectively,
{${\mathchoice{\mbox{\boldmath$\displaystyle H$}}                      {\mbox{\boldmath$\textstyle H$}}                      {\mbox{\boldmath$\scriptstyle H$}}                      {\mbox{\boldmath$\scriptscriptstyle H$}}}%
$ is a diagonal matrix of the channel frequency response},
${\mathchoice{\mbox{\boldmath$\displaystyle D$}}                      {\mbox{\boldmath$\textstyle D$}}                      {\mbox{\boldmath$\scriptstyle D$}}                      {\mbox{\boldmath$\scriptscriptstyle D$}}}%
$ is $N\times1$ clipping noise vector, and
${\mathchoice{\mbox{\boldmath$\displaystyle Z$}}                      {\mbox{\boldmath$\textstyle Z$}}                      {\mbox{\boldmath$\scriptstyle Z$}}                      {\mbox{\boldmath$\scriptscriptstyle Z$}}}%
$ is AWGN vector. {Starting from (\ref{Alsa1}), we equalize the channel by
multiplying with
${\mathchoice{\mbox{\boldmath$\displaystyle H$}}                      {\mbox{\boldmath$\textstyle H$}}                      {\mbox{\boldmath$\scriptstyle H$}}                      {\mbox{\boldmath$\scriptscriptstyle H$}}}%
^{-1}$ and select the rows whose indices correspond to locations of the
reserved tones by multiplying with a $M\times N$ row selection matrix
${\mathchoice{\mbox{\boldmath$\displaystyle S$}}                      {\mbox{\boldmath$\textstyle S$}}                      {\mbox{\boldmath$\scriptstyle S$}}                      {\mbox{\boldmath$\scriptscriptstyle S$}}}%
_{r}$. This results in}
\begin{equation}
\underbrace
{{\mathchoice{\mbox{\boldmath$\displaystyle S$}} {\mbox{\boldmath$\textstyle S$}} {\mbox{\boldmath$\scriptstyle S$}} {\mbox{\boldmath$\scriptscriptstyle S$}}}%
_{r}%
{\mathchoice{\mbox{\boldmath$\displaystyle H$}} {\mbox{\boldmath$\textstyle H$}} {\mbox{\boldmath$\scriptstyle H$}} {\mbox{\boldmath$\scriptscriptstyle H$}}}%
^{-1}%
{\mathchoice{\mbox{\boldmath$\displaystyle Y$}} {\mbox{\boldmath$\textstyle Y$}} {\mbox{\boldmath$\scriptstyle Y$}} {\mbox{\boldmath$\scriptscriptstyle Y$}}}%
}%
_{{\mathchoice{\mbox{\boldmath$\displaystyle g$}} {\mbox{\boldmath$\textstyle g$}} {\mbox{\boldmath$\scriptstyle g$}} {\mbox{\boldmath$\scriptscriptstyle g$}}}%
}=\underbrace
{{\mathchoice{\mbox{\boldmath$\displaystyle S$}} {\mbox{\boldmath$\textstyle S$}} {\mbox{\boldmath$\scriptstyle S$}} {\mbox{\boldmath$\scriptscriptstyle S$}}}%
_{r}%
{\mathchoice{\mbox{\boldmath$\displaystyle F$}} {\mbox{\boldmath$\textstyle F$}} {\mbox{\boldmath$\scriptstyle F$}} {\mbox{\boldmath$\scriptscriptstyle F$}}}%
}%
_{{\mathchoice{\mbox{\boldmath$\displaystyle \Phi$}} {\mbox{\boldmath$\textstyle \Phi$}} {\mbox{\boldmath$\scriptstyle \Phi$}} {\mbox{\boldmath$\scriptscriptstyle \Phi$}}}%
}\underbrace
{{\mathchoice{\mbox{\boldmath$\displaystyle d$}} {\mbox{\boldmath$\textstyle d$}} {\mbox{\boldmath$\scriptstyle d$}} {\mbox{\boldmath$\scriptscriptstyle d$}}}%
}%
_{{\mathchoice{\mbox{\boldmath$\displaystyle f$}} {\mbox{\boldmath$\textstyle f$}} {\mbox{\boldmath$\scriptstyle f$}} {\mbox{\boldmath$\scriptscriptstyle f$}}}%
}%
+{\mathchoice{\mbox{\boldmath$\displaystyle S$}} {\mbox{\boldmath$\textstyle S$}} {\mbox{\boldmath$\scriptstyle S$}} {\mbox{\boldmath$\scriptscriptstyle S$}}}%
_{r}%
{\mathchoice{\mbox{\boldmath$\displaystyle H$}} {\mbox{\boldmath$\textstyle H$}} {\mbox{\boldmath$\scriptstyle H$}} {\mbox{\boldmath$\scriptscriptstyle H$}}}%
^{-1}%
{\mathchoice{\mbox{\boldmath$\displaystyle Z$}} {\mbox{\boldmath$\textstyle Z$}} {\mbox{\boldmath$\scriptstyle Z$}} {\mbox{\boldmath$\scriptscriptstyle Z$}}}%
, \label{Alsa2}%
\end{equation}
where
${\mathchoice{\mbox{\boldmath$\displaystyle F$}}                      {\mbox{\boldmath$\textstyle F$}}                      {\mbox{\boldmath$\scriptstyle F$}}                      {\mbox{\boldmath$\scriptscriptstyle F$}}}%
$ is the DFT matrix and
${\mathchoice{\mbox{\boldmath$\displaystyle D$}}                      {\mbox{\boldmath$\textstyle D$}}                      {\mbox{\boldmath$\scriptstyle D$}}                      {\mbox{\boldmath$\scriptscriptstyle D$}}}%
={\mathchoice{\mbox{\boldmath$\displaystyle F$}}                      {\mbox{\boldmath$\textstyle F$}}                      {\mbox{\boldmath$\scriptstyle F$}}                      {\mbox{\boldmath$\scriptscriptstyle F$}}}%
{\mathchoice{\mbox{\boldmath$\displaystyle d$}}                      {\mbox{\boldmath$\textstyle d$}}                      {\mbox{\boldmath$\scriptstyle d$}}                      {\mbox{\boldmath$\scriptscriptstyle d$}}}%
$. As seen in (\ref{Alsa2}), the clipping noise on the reserved tones is used
as measurements to recover the clipping noise
${\mathchoice{\mbox{\boldmath$\displaystyle d$}}                      {\mbox{\boldmath$\textstyle d$}}                      {\mbox{\boldmath$\scriptstyle d$}}                      {\mbox{\boldmath$\scriptscriptstyle d$}}}%
$ in time domain by sparse recovery algorithm. Additionally, in
\cite{Al-Safadi2}, a method exploiting a-priori information together with
weighted $l_{1}$ minimization for enhanced recovery followed by Bayesian
techniques is proposed. However, the performance of \cite{Al-Safadi} and
\cite{Al-Safadi2} is restricted due to weakness of the compressed sensing
against noise.

In \cite{Kim}, more enhanced clipping noise cancellation scheme using
compressed sensing is proposed. Different from \cite{Al-Safadi} and
\cite{Al-Safadi2}, the scheme in \cite{Kim} does not cause data rate loss,
because it exploits the clipping noise in frequency domain as measurements
underlying in the data tones rather than the reserved tones. In this case,
transmitted data and clipping noise are mixed in the data tones. To
distinguish the clipping noise from the data tones well, this scheme exploits
part of the received data tones with high reliability. To (\ref{Alsa1}), we
multiply
${\mathchoice{\mbox{\boldmath$\displaystyle H$}}                      {\mbox{\boldmath$\textstyle H$}}                      {\mbox{\boldmath$\scriptstyle H$}}                      {\mbox{\boldmath$\scriptscriptstyle H$}}}%
^{-1}$ and row selection matrix
${\mathchoice{\mbox{\boldmath$\displaystyle S$}}                      {\mbox{\boldmath$\textstyle S$}}                      {\mbox{\boldmath$\scriptstyle S$}}                      {\mbox{\boldmath$\scriptscriptstyle S$}}}%
_{d}$, selecting the locations of reliable data tones, as
\begin{equation}
{\mathchoice{\mbox{\boldmath$\displaystyle S$}} {\mbox{\boldmath$\textstyle S$}} {\mbox{\boldmath$\scriptstyle S$}} {\mbox{\boldmath$\scriptscriptstyle S$}}}%
_{d}%
{\mathchoice{\mbox{\boldmath$\displaystyle H$}} {\mbox{\boldmath$\textstyle H$}} {\mbox{\boldmath$\scriptstyle H$}} {\mbox{\boldmath$\scriptscriptstyle H$}}}%
^{-1}%
{\mathchoice{\mbox{\boldmath$\displaystyle Y$}} {\mbox{\boldmath$\textstyle Y$}} {\mbox{\boldmath$\scriptstyle Y$}} {\mbox{\boldmath$\scriptscriptstyle Y$}}}%
={\mathchoice{\mbox{\boldmath$\displaystyle S$}} {\mbox{\boldmath$\textstyle S$}} {\mbox{\boldmath$\scriptstyle S$}} {\mbox{\boldmath$\scriptscriptstyle S$}}}%
_{d}%
{\mathchoice{\mbox{\boldmath$\displaystyle C$}} {\mbox{\boldmath$\textstyle C$}} {\mbox{\boldmath$\scriptstyle C$}} {\mbox{\boldmath$\scriptscriptstyle X$}}}%
+{\mathchoice{\mbox{\boldmath$\displaystyle S$}} {\mbox{\boldmath$\textstyle S$}} {\mbox{\boldmath$\scriptstyle S$}} {\mbox{\boldmath$\scriptscriptstyle S$}}}%
_{d}%
{\mathchoice{\mbox{\boldmath$\displaystyle D$}} {\mbox{\boldmath$\textstyle D$}} {\mbox{\boldmath$\scriptstyle D$}} {\mbox{\boldmath$\scriptscriptstyle D$}}}%
+{\mathchoice{\mbox{\boldmath$\displaystyle S$}} {\mbox{\boldmath$\textstyle S$}} {\mbox{\boldmath$\scriptstyle S$}} {\mbox{\boldmath$\scriptscriptstyle S$}}}%
_{d}%
{\mathchoice{\mbox{\boldmath$\displaystyle H$}} {\mbox{\boldmath$\textstyle H$}} {\mbox{\boldmath$\scriptstyle H$}} {\mbox{\boldmath$\scriptscriptstyle H$}}}%
^{-1}%
{\mathchoice{\mbox{\boldmath$\displaystyle Z$}} {\mbox{\boldmath$\textstyle Z$}} {\mbox{\boldmath$\scriptstyle Z$}} {\mbox{\boldmath$\scriptscriptstyle Z$}}}%
. \label{Kim1}%
\end{equation}
Then, we estimate the
${\mathchoice{\mbox{\boldmath$\displaystyle S$}}                      {\mbox{\boldmath$\textstyle S$}}                      {\mbox{\boldmath$\scriptstyle S$}}                      {\mbox{\boldmath$\scriptscriptstyle S$}}}%
_{d}%
{\mathchoice{\mbox{\boldmath$\displaystyle \hat{C}$}}                      {\mbox{\boldmath$\textstyle \hat{C}$}}                      {\mbox{\boldmath$\scriptstyle \hat{C}$}}                      {\mbox{\boldmath$\scriptscriptstyle \hat{C}$}}}%
$ and subtract them from (\ref{Kim1}) as
\begin{equation}
\underbrace
{{\mathchoice{\mbox{\boldmath$\displaystyle S$}} {\mbox{\boldmath$\textstyle S$}} {\mbox{\boldmath$\scriptstyle S$}} {\mbox{\boldmath$\scriptscriptstyle S$}}}%
_{d}%
{\mathchoice{\mbox{\boldmath$\displaystyle H$}} {\mbox{\boldmath$\textstyle H$}} {\mbox{\boldmath$\scriptstyle H$}} {\mbox{\boldmath$\scriptscriptstyle H$}}}%
^{-1}%
{\mathchoice{\mbox{\boldmath$\displaystyle Y$}} {\mbox{\boldmath$\textstyle Y$}} {\mbox{\boldmath$\scriptstyle Y$}} {\mbox{\boldmath$\scriptscriptstyle Y$}}}%
-{\mathchoice{\mbox{\boldmath$\displaystyle S$}} {\mbox{\boldmath$\textstyle S$}} {\mbox{\boldmath$\scriptstyle S$}} {\mbox{\boldmath$\scriptscriptstyle S$}}}%
_{d}%
{\mathchoice{\mbox{\boldmath$\displaystyle \hat{C}$}} {\mbox{\boldmath$\textstyle \hat{C}$}} {\mbox{\boldmath$\scriptstyle \hat{C}$}} {\mbox{\boldmath$\scriptscriptstyle \hat{C}$}}}%
}%
_{{\mathchoice{\mbox{\boldmath$\displaystyle g$}} {\mbox{\boldmath$\textstyle g$}} {\mbox{\boldmath$\scriptstyle g$}} {\mbox{\boldmath$\scriptscriptstyle g$}}}%
}=\underbrace
{{\mathchoice{\mbox{\boldmath$\displaystyle S$}} {\mbox{\boldmath$\textstyle S$}} {\mbox{\boldmath$\scriptstyle S$}} {\mbox{\boldmath$\scriptscriptstyle S$}}}%
_{d}%
{\mathchoice{\mbox{\boldmath$\displaystyle F$}} {\mbox{\boldmath$\textstyle F$}} {\mbox{\boldmath$\scriptstyle F$}} {\mbox{\boldmath$\scriptscriptstyle F$}}}%
}%
_{{\mathchoice{\mbox{\boldmath$\displaystyle \Phi$}} {\mbox{\boldmath$\textstyle \Phi$}} {\mbox{\boldmath$\scriptstyle \Phi$}} {\mbox{\boldmath$\scriptscriptstyle \Phi$}}}%
}\underbrace
{{\mathchoice{\mbox{\boldmath$\displaystyle d$}} {\mbox{\boldmath$\textstyle d$}} {\mbox{\boldmath$\scriptstyle d$}} {\mbox{\boldmath$\scriptscriptstyle d$}}}%
}%
_{{\mathchoice{\mbox{\boldmath$\displaystyle f$}} {\mbox{\boldmath$\textstyle f$}} {\mbox{\boldmath$\scriptstyle f$}} {\mbox{\boldmath$\scriptscriptstyle f$}}}%
}%
+{\mathchoice{\mbox{\boldmath$\displaystyle S$}} {\mbox{\boldmath$\textstyle S$}} {\mbox{\boldmath$\scriptstyle S$}} {\mbox{\boldmath$\scriptscriptstyle S$}}}%
_{d}%
({\mathchoice{\mbox{\boldmath$\displaystyle C$}} {\mbox{\boldmath$\textstyle C$}} {\mbox{\boldmath$\scriptstyle C$}} {\mbox{\boldmath$\scriptscriptstyle C$}}}%
-{\mathchoice{\mbox{\boldmath$\displaystyle \hat{C}$}} {\mbox{\boldmath$\textstyle \hat{C}$}} {\mbox{\boldmath$\scriptstyle \hat{C}$}} {\mbox{\boldmath$\scriptscriptstyle \hat{C}$}})}%
+{\mathchoice{\mbox{\boldmath$\displaystyle S$}} {\mbox{\boldmath$\textstyle S$}} {\mbox{\boldmath$\scriptstyle S$}} {\mbox{\boldmath$\scriptscriptstyle S$}}}%
_{d}%
{\mathchoice{\mbox{\boldmath$\displaystyle H$}} {\mbox{\boldmath$\textstyle H$}} {\mbox{\boldmath$\scriptstyle H$}} {\mbox{\boldmath$\scriptscriptstyle H$}}}%
^{-1}%
{\mathchoice{\mbox{\boldmath$\displaystyle Z$}} {\mbox{\boldmath$\textstyle Z$}} {\mbox{\boldmath$\scriptstyle Z$}} {\mbox{\boldmath$\scriptscriptstyle Z$}}}%
. \label{Kim2}%
\end{equation}
Then, from partially extracted clipping noise in frequency domain, we can
recover the clipping noise
${\mathchoice{\mbox{\boldmath$\displaystyle d$}}                      {\mbox{\boldmath$\textstyle d$}}                      {\mbox{\boldmath$\scriptstyle d$}}                      {\mbox{\boldmath$\scriptscriptstyle d$}}}%
$ in time domain via sparse recovery algorithms. Furthermore, this scheme can
adjust the number of the measurements $M$ by changing the reliability of
received data. Therefore, when there is AWGN noise, we can select the optimal
number of measurements corresponding to the noise amount. Consequently, this
scheme successfully realizes the clipping noise cancellation scheme by
overcoming weakness of the compressed sensing against noise. Additionally, in
\cite{Kim}, clipping noise cancellation for orthogonal frequency-division
multiple access (OFDMA) systems is also proposed using compressed sensing. The
fast Fourier transform (FFT) block of OFDM systems can be decomposed into the
small FFT blocks. And, the subset of rows in the small sized DFT matrix can
also be used as a sensing matrix, which can be used to recover the clipping
noise for OFDMA systems via sparse recovery algorithm.

\begin{figure}[t]
\centering\includegraphics[width=0.6\columnwidth,height=5.9cm]{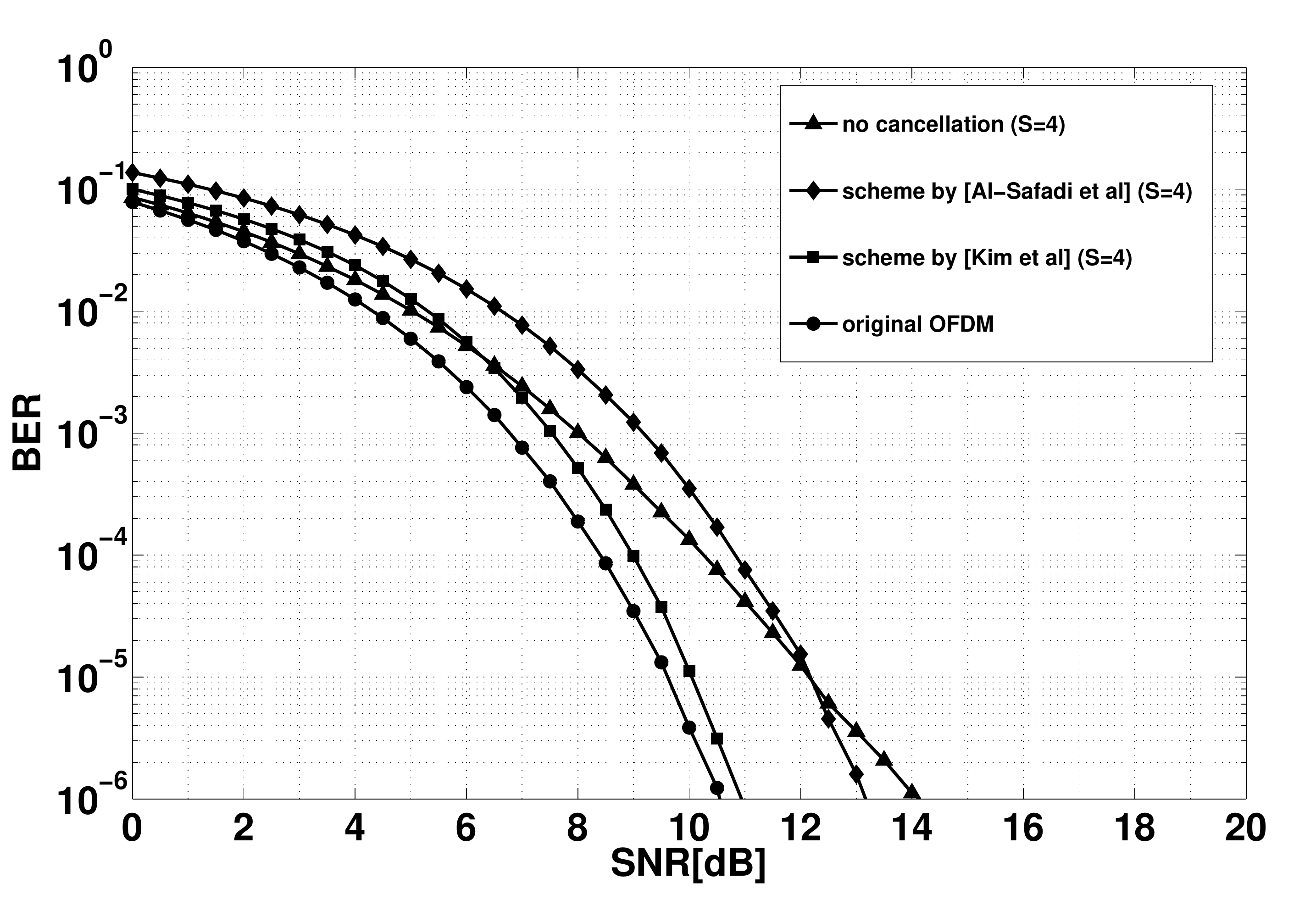}\caption{BER
performance in OFDM AWGN channel using compressed sensing based clipping noise
cancellation schemes as described in Al-Safadi et al \cite{Al-Safadi} and Kim
et al \cite{Kim}. In \cite{Kim} 23 out of 64 QPSK subcarriers are used for
recovery of the $S=4$ sparse clipping noise. In \cite{Al-Safadi} 29 tones are
selected by a (59, 29, 14) difference set and reserved. Baseline performance
is the original OFDM signal with no clipping. It can be seen that provided the
AWGN is not too strong, sparsity in the clipping noise can lead to significant
performance gains of more than 3 dB given an error probability of $10^{-6}$.}%
\label{fig_cs}%
\end{figure}

Fig. \ref{fig_cs} shows the bit error rate (BER) over signal-to-noise ratio
(SNR) performance of the clipping noise cancellation schemes based on
compressed sensing described in \cite{Al-Safadi} and \cite{Kim} for OFDM
signals over the AWGN channel. The $S$-sparse clipping noise signal
contaminates the original OFDM signal and the case of no clipping noise
cancellation shows the worst BER performance among all schemes. In
\cite{Al-Safadi}, the authors applied the compressed sensing technique to OFDM
systems for the first time, but there is a benefit only for the high SNR
region due to weakness of compressed sensing recovery against AWGN. The BER
performance of the scheme in \cite{Kim} is better because the number of the
measurements can be adjusted corresponding to the AWGN level.

\section{Conclusions}

{Despite two decades of intensive research the PAPR problem remains one of the
major problems in multicarrier theory with huge practical impact. This article
provides a fresh look on this problem by outlining a new perspective using
alternative metrics (including MIMO and multiuser systems as a special case),
the corresponding theoretical foundations and related designs. This is
followed by thorough discussion of current limits and new future directions.}

\section{Acknowledgements}

The work of G. Wunder was supported by the Deutsche Forschungsgemeinschaft
(DFG) under grant WU 598/3-1.

The work of J.-S. No was supported by the National Research Foundation of
Korea (NRF) grant funded by the Korea government (MEST) (No. 2012-0000186).

\bibliographystyle{IEEEbib}
\bibliography{spm_literature}

\section{Biographies}

\textbf{Gerhard Wunder} (M'05) Gerhard Wunder (M'05) studied electrical engineering at the University of
Hannover, Germany, and the Technische Universit\"{a}t (TU) Berlin, Germany,
and received his graduate degree in electrical engineering (Dipl.-Ing.) with
highest honors in 1999 and the PhD degree (Dr.-Ing.) in communication
engineering on the peak-to-average power ratio (PAPR) problem in OFDM with
distinction (summa cum laude) in 2003 from TU Berlin. In 2007, he also
received the habilitation degree (venia legendi) and became a Privatdozent at
the TU Berlin in the field of detection/estimation theory, stochastic
processes and information theory. Since 2003 he is heading a research group at
the Fraunhofer Lab for Mobile Communications (FhG-MCI), Heinrich-Hertz-Institut, working in close
collaboration with industry on theoretical and practical problems in wireless
communication networks particularly in the field of LTE-A systems. He is a
recipient of research fellowships from the German national research foundation.

In 2000 and 2005, he was a visiting professor at the Georgia Institute of
Technology (Prof. Jayant) in Atlanta (USA, GA), and the Stanford University
(Prof. Paulraj) in Palo Alto/USA (CA). In 2009 he was a consultant at
Alcatel-Lucent Bell Labs (USA, NJ), both in Murray Hill and Crawford Hill. He
was a general co-chair of the 2009 International ITG Workshop on Smart
Antennas (WSA 2009) and a lead guest editor in 2011 for a special issue of the
Journal of Advances on Signal Processing regarding the PAPR problem of the
European Association for Signal Processing. Since 2011, he is also an editor
for the IEEE Transactions on Wireless Communications (TWireless) in the area
of Wireless Communications Theory and Systems (WCTS). In 2011 Dr. Wunder
received the best paper award for outstanding scientific publication in the
field of communication engineering by the German communication engineering
society (ITG Award 2011).

\textbf{Robert F. H. Fischer} (M'99-SM'10) received the Dr.-Ing. degree in 1996, and the habilitation
degree in 2001, from the University of Erlangen--N{\"u}rnberg, Erlangen,
Germany. The subject of his dissertation was multichannel and multicarrier
modulation, that of his habilitation was precoding and signal shaping. Form
1992 to 1996, he was a Research Assistant at the Telecommunications Institute,
University of Erlangen--N{\"u}rnberg. During 1997, he was with the IBM
Research Laboratory, Z{\"u}rich, Switzerland. In 1998, he returned to the
Telecommunications Institute II, University of Erlangen--N{\"u}rnberg, and in
2005 he spent a sabbatical at ETH, Z{\"u}rich, Switzerland. Since 2011, he has
been full professor at the University of Ulm, Germany. Currently, he teaches
undergraduate and graduate courses on signals and systems and on digital
communications. His research concentrates on fast digital transmission
including single- and multicarrier modulation techniques. His current
interests are information theory, coded modulation, digital communications and
signal processing, and especially precoding and shaping techniques for
high-rate transmission schemes. Dr. Fischer received the Dissertation Award
form the Technische Fakult{\"a}t, University of Erlangen--N{\"u}rnberg, in
1997, the Publication Award of the German Society of Information Techniques
(ITG) in 2000, the Wolfgang Finkelnburg habilitation award in 2002, and the
Philipp-Reis-Preis in 2005. He is author of the textbook ``Precoding and
Signal Shaping for Digital Transmission'' (John Wiley \& Sons, New York, 2002).

\textbf{Holger Boche} (M'04-SM'07-F'11) received the Dipl.-Ing. and Dr.-Ing. degrees in
electrical engineering from the Technische Universitaet Dresden, Dresden,
Germany, in 1990 and 1994, respectively. He graduated in mathematics from the
Technische Universitaet Dresden in 1992. From 1994 to 1997, he did
postgraduate studies in mathematics at the Friedrich-Schiller Universit{{\"a}%
}t Jena, Jena, Germany. He received his Dr.Rer.Nat. degree in pure mathematics
from the Technische Universitaet Berlin, Berlin, Germany, in 1998. In 1997, he
joined the Heinrich-Hertz-Institut (HHI) f{{\"u}}r Nachrichtentechnik Berlin,
Berlin, Germany. Since 2002, he has been a Full Professor for mobile
communication networks with the Institute for Communications Systems,
Technische Universit{{\"a}}t Berlin. In 2003, he became Director of the
Fraunhofer German-Sino Lab for Mobile Communications, Berlin, Germany, and
since 2004 he has also been Director of the Fraunhofer Institute for
Telecommunications (HHI), Berlin, Germany. Since, October 2010 he is with the
Institute of Theoretical Information Technology and Full Professor at the
Technical University of Munich, Munich, Germany. He was a Visiting Professor
with the ETH Zurich, Zurich, Switzerland, during the 2004 and 2006 Winter
terms, and with KTH Stockholm, Stockholm, Sweden, during the 2005 Summer term.
Prof. Boche is a Member of IEEE Signal Processing Society SPCOM and SPTM
Technical Committee. He was elected a Member of the German Academy of Sciences
(Leopoldina) in 2008 and of the Berlin Brandenburg Academy of Sciences and
Humanities in 2009. He received the Research Award \textquotedblleft
Technische Kommunikation\textquotedblright\ from the Alcatel SEL Foundation in
October 2003, the \textquotedblleft Innovation Award\textquotedblright\ from
the Vodafone Foundation in June 2006, and the Gottfried Wilhelm Leibniz Prize
from the Deutsche Forschungsgemeinschaft (German Research Foundation) in 2008.
He was co-recipient of the 2006 IEEE Signal Processing Society Best Paper
Award and recipient of the 2007 IEEE Signal Processing Society Best Paper Award.

\textbf{Simon Litsyn} (M'94-SM'99) was born in
Khar'kov, U.S.S.R., in 1957. He received the M.Sc. degree from Perm
Polytechnical Institute, Perm, U.S.S.R., in 1979 and the Ph.D.degree from
Leningrad Electrotechnical Institute, Leningrad, U.S.S.R., in 1982, all in
electrical engineering. Since 1991, he has been with the Department of
Electrical Engineering-Systems, Tel-Aviv University, Tel-Aviv, Israel, where
he is a Professor. Since 2005 he also works in Sandisk, where he is Chief
Scientist. His research interests include coding theory, communications and
applications of discrete mathematics. He authored "Covering Codes", Elsevier,
1997, and "Peak Power Control in Multicarrier Communications", Cambridge
University Press, 2007. Dr. Litsyn received the Guastallo Fellowship in 1992.
In 2000-2003 he has served as an Associate Editor for Coding Theory for IEEE
Transactions on Information Theory. He is an editor of Advances in Mathematics
of Communications (AMC) and Applicable Algebra in Engineering, Communications
and Computing (AAECC) journals.

\textbf{Jong-Seon No} (M'82-SM'10-F'12) received the B.S. and M.S.E.E. degrees in Electronics
Engineering from Seoul National University, Seoul, Korea, in 1981 and 1984,
respectively, and the Ph.D. degree in Electrical Engineering from the
University of Southern California, Los Angeles, in 1988. He was a Senior MTS
with Hughes Network Systems, Germantown, MD, from February 1988 to July 1990.
He was an Associate Professor with the Department of Electronic Engineering,
Konkuk University, Seoul, from September 1990 to July 1999. He joined the
Faculty of the Department of Electrical Engineering and Computer Science,
Seoul National University, in August 1999, where he is currently a Professor.
He served as a General Co-Chair for International Symposium on Information
Theory and Its Applications 2006 (ISITA 2006), Seoul, Korea hosted by IEICE
and International Symposium on Information Theory 2009 (ISIT 2009), Seoul,
Korea hosted by IEEE Information Theory Society. His research interests
include error-correcting codes, sequences, cryptography, space-time codes,
LDPC codes, network coding, compressed sensing, and wireless communication systems.

\end{document}